\documentclass[%
twocolumn,
superscriptaddress,
groupedaddress,
 amsmath,amssymb,
]{revtex4-2}

\usepackage{graphicx}% Include figure files
\usepackage{dcolumn}% Align table columns on decimal point
\usepackage[normalem]{ulem} % for \st (strikeout) command
\usepackage{bm}% bold math
\usepackage{hyperref}% add hypertext capabilities
\usepackage{soul,xcolor}
\usepackage[utf8]{inputenc}
\usepackage{textcomp}
\DeclareUnicodeCharacter{030A}{\r{}}

\hypersetup{
    colorlinks=true,
    linkcolor=blue,
    filecolor=magenta,      
    urlcolor=cyan,
    pdftitle={Sharelatex Example},
    bookmarks=true,
    citecolor=blue
}

\begin{document}

\preprint{APS/123-QED}

\title{Spatially-Controlled Planar Guided Crystallization of Low-Loss Phase Change Materials for Programmable Photonics}

\author{Fouad Bentata$^{1,2,3}$}%
\author{Arnaud Taute$^1$}%
\author{Capucine Laprais$^1$}%
\author{Régis Orobtchouk$^1$}%
\author{Eva Kempf$^3$}%
\author{Alban Gassenq$^4$}%
\author{Yves Pipon$^5$}%
\author{Michele Calvo$^3$}%
\author{Valérie Martinez$^4$}%
\author{Stéphane Monfray$^3$}%
\author{Guillaume Saint-Girons$^1$}%
\author{Nicolas Baboux$^1$}%
\author{Hai Son Nguyen$^1$}%
\author{Xavier Letartre$^1$}%
\author{Lotfi Berguiga$^1$}%
\author{Patrice Genevet$^{2,6}$}%
\author{Sébastien Cueff$^1$}%
\email{sebastien.cueff@cnrs.fr}

\affiliation{$^1$CNRS, Ecole Centrale de Lyon, INSA Lyon, Universite Claude Bernard Lyon 1, CPE Lyon, INL, UMR5270, 69130 Ecully, France}%
\affiliation {$^2$Université Côte d'Azur, CNRS, CRHEA, Rue Bernard Gregory, Valbonne, 06560, France.}
\affiliation{$^3$STMicroelectronics, 850 Rue Jean Monnet, Crolles, 38920, France}
\affiliation{$^4$Universite Claude Bernard Lyon 1, CNRS, Institut Lumière Matière, UMR5306, F-69100, Villeurbanne, France}
\affiliation{$^5$Univ. Claude Bernard Lyon 1 - CNRS/IN2P3, IP2I (UMR 5822), 69622 Villeurbanne, France}
\affiliation{$^6$Department of Physics, Colorado School of Mines, 1523 Illinois Street, Golden, CO, 80401, USA}


\begin{abstract}
Photonic integrated devices are progressively evolving beyond passive components into fully programmable systems, notably driven by the progress in chalcogenide phase-change materials (PCMs)  for non-volatile reconfigurable nanophotonics. However, the stochastic nature of their crystal grain formation results in strong spatial and temporal crystalline inhomogeneities. Here, we propose the concept of spatially-controlled planar guided crystallization, a novel method for programming the growth of optically homogeneous low-loss Sb$_2$S$_3$ PCM, leveraging the seeded directional and progressive crystallization within confined channels. This guided crystallization method is experimentally shown to circumvent the current limitations of conventional PCM-based nanophotonic devices, including a multilevel non-volatile optical phase-shifter exploiting a silicon nitride-based Mach-Zehnder interferometer, and a programmable metasurface with spectrally reconfigurable bound state in the continuum. Precisely controlling the growth of PCMs to ensure optically uniform crystalline properties across devices is the cornerstone for the industrial development of non-volatile reconfigurable photonic integrated circuits.
\end{abstract}

\keywords{Phase-change material, Phase-shifter, Mach-Zehnder Interferometer, Silicon Nitride}

\maketitle

%%%%%%%%%%%%%%%%%%%%%%%%%%  body  %%%%%%%%%%%%%%%%%%%%%%%%%%
\section{Introduction}

Integrating designer nanostructures on a chip opens up unprecedented functionalities for microelectronics and photonics, for various applications ranging from light-speed data communications and calculations to beam shaping and focusing with optical metasurfaces \cite{Genevet_2015,yu2014flat,shastri2021photonics,wang2024advances,cheben2018subwavelength, VCSEL_2020}. Passive photonic integrated circuits have now reached a high degree of sophistication, providing full control of nearly all aspects of light at the nanoscale, and can now readily be included in complex microelectronic chips. Research on large-scale systems for optical datacoms, metalenses, and neuromorphic computing is now close to leaping industrialization but requires tunable building blocks \cite{mikheeva2022space,kuznetsov2024roadmap,brongersma2025second}.

CMOS-compatible solutions to locally tune devices do exist in foundries, for example, electrically-driven phase-shifters leveraging the thermo-optic effect or plasma-dispersion effect in silicon-based waveguides \cite{bogaerts2020programmable,liu2016fully,perez2017multipurpose}. Using this technology, impressive results were obtained on photonic-based deep learning and optical FPGA in the general framework of programmable photonics. Other efficient solutions include micro-electro-mechanical systems (MEMS) and liquid crystals \cite{li2019phase,van2022low,feng2020performance}. While these different technologies enabled proof of principles for novel functionalities in integrated photonics, they rely on volatile modulation schemes: their operations require a constant supply of electrical power to operate. Such platforms may, therefore, not be scalable and sustainable for future systems incorporating several thousands of building blocks.

Phase change materials (PCMs) have recently become a popular class of materials to actively tune integrated photonic devices \cite{wuttig2017phase,abdollahramezani2020tunable}. PCMs such as Ge$_2$Sb$_2$Te$_5$ (GST) are exploited for their unique physical properties, and most notably their ability to rapidly and reversibly switch between crystalline and amorphous phases exhibiting widely different optical and electrical properties. For example, the tunability of their resistivity is industrially developed to create phase change memories \cite{raoux2008phase}. On the other hand, the dynamic modulation of their optical properties, i.e. refractive index and extinction coefficient, are exciting opportunities to create integrated amplitude modulators and switches in waveguides or metasurfaces \cite{gholipour2013all,rios2015integrated,cheng2018device, howes2020optical,tripathi2021tunable, cueff2020vo2, cueff2021reconfigurable, li2016reversible, wang2021electrical, zhang2025chip}. The non-volatile nature of this refractive index change is a striking advantage over conventional modulators since maintaining continuous power is no longer required for phase programming.

 The current roadblock that limits their applicability for photonics is their high optical absorption in the near-infrared region. Even though some lower-loss alternatives to GST, such as GeSbSeTe have been reported \cite{zhang2019broadband}, the level of optical absorption may still be too high for applications such as phase modulation in large-scale photonic integrated circuits requiring a large refractive index change and transparency in the near-infrared region.

In this context, Sb$_2$S$_3$ and Sb$_2$Se$_3$ represent a very promising class of emerging PCMs with experimentally demonstrated large refractive index changes upon crystallization and ultra-low extinction coefficients from the visible range to infrared \cite{dong2019wide,delaney2020new,bieganski2024sb}. Recent works exploited these low-loss PCMs to control the optical properties of integrated photonic devices such as micro-rings, Mach-Zehnder interferometers (MZI), metasurfaces and multimode interferometers (MMI) using Sb$_2$S$_3$ \cite{fang2021non, chen2023non, fang2024nonvolatile, moitra2023programmable} and Sb$_2$Se$_3$ \cite{delaney2021nonvolatile, rios2022ultra, fang2022ultra, yang2023non, tara2024non, wu2024freeform}. 

However, the stochastic nature of nucleation and crystal grain formation in Sb$_2$S$_3$ and Sb$_2$Se$_3$ give rise to spatially inhomogeneous crystallization, resulting in polycrystalline layers. Furthermore, this polycristallinity combined with the orthorhombic crystalline structure of Sb$_2$S$_3$ lead to uncontrolled optical anisotropy \cite{laprais2024reversible}. The situation gets even more problematic at the micro-nano-scale: given the random spatial distribution of crystals of Sb$_2$S$_3$, the energy and time requirements to bring one specific Sb$_2$S$_3$-based device to a precise intermediate state will very likely vary from one device to another. Since continuous control of intermediate states is key for future tunable photonic devices and non Von-Neuman computing schemes, deep understanding and precise manipulation of crystal spatial arrangements in these emerging PCMs is crucial.

Here, we propose and demonstrate a novel method enabling spatially-controlled crystallization of PCMs, using an original guided planar growth scheme. By exploiting both the isothermal time-crystallization \cite{taute2023programming} and guided crystallization scheme in Sb$_2$S$_3$, we show how to precisely control the spatial flow of crystallization, similar to a fluid in a channel. We demonstrate the exceptional potential of this guided crystallization for advancing both guided optics and free-space optics. We first present a multilevel non-volatile phase shifter in the near-infrared region using a hybrid SiN/Sb$_2$S$_3$ Mach-Zehnder interferometer. We demonstrate how this guided crystallization enables precise control over the temporal and spatial evolution of crystals on the chip, as well as their overall quality. By strategically designing crystallization reservoirs and channels, we achieve in-plane growth of crystals with homogeneous optical properties, analogous to a planar and spatially-controlled version of liquid-to-solid crystal growth schemes such as directional solidification or Czochralski growth \cite{dai2008grain,raza2019grain,uecker2014historical}. Furthermore, we show that this controlled in-plane growth significantly enhances device performance, and we prove this concept experimentally by producing a reconfigurable bound state in the continuum (BIC) metasurface with an extremely wide and controlled $100~nm$ spectral modulation.

 \section{Results}
\subsection{Spatially-controlled planar crystal growth concept}

\begin{figure*}[htbp!]
\centering\includegraphics[width=1\linewidth]{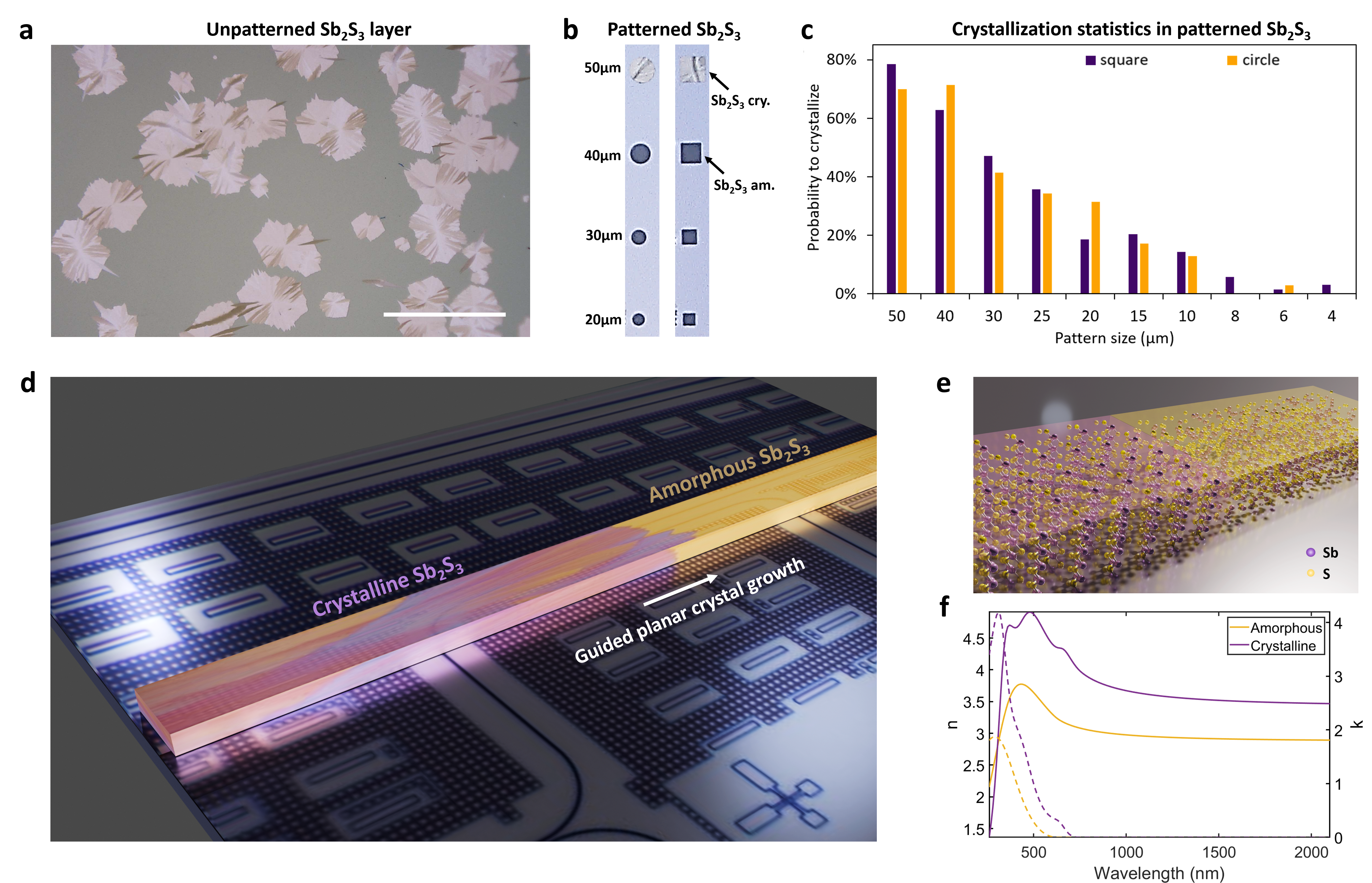}
\caption{\textbf{General working principle of spatially-controlled planar growth of Sb$_2$S$_3$.} a) Optical microscope image of an Sb$_2$S$_3$ thin-film at the early stage of crystallization at 260°C. The crystallization occurs via stochastic nucleation, followed by the growth of micro-crystals (light areas) around nuclei in the amorphous host (grey areas). (Scale bar: 200 µm). b) Optical microscope image of micro-patterned 42-nm-thick Sb$_2$S$_3$ (squares and circles) subjected to thermal heating at 280°C for 2 hours. The appearance of nuclei being stochastic, confining Sb$_2$S$_3$ in small volumes drastically reduces the probability of crystallization. c) Statistics of the crystallization of patterned Sb$_2$S$_3$ for different dimensions. Below 10 $\mu$m, the probability for crystallization approaches zero: the microstructures remain amorphous even at high temperatures and long duration. d) Schematic of the proposed guided crystallization concept: the PCM is patterned as a channel whose volume is just enough to contain one nucleus. Heating it to 250 °C, the nucleus acts as a seed for the crystal growth, allowing the initially amorphous Sb$_2$S$_3$ to progressively crystallize along the channel. The crystallization is spatially controlled by the pattern and time-controlled by the heating. e) Microscopic rendering of the moving boundary between crystalline (magenta) and amorphous (yellow) Sb$_2$S$_3$ in the guiding channel; f) Complex refractive index of the amorphous (yellow) and crystalline (violet) phase of Sb$_2$S$_3$, showing a large modulation of refractive index (solid lines) upon crystallization and a negligible extinction coefficient (dashed lines) in both phases. } 
\label{fig:GuidedCrystallization}
\end{figure*}

Emerging low-loss PCMs such as Sb$_2$S$_3$ are known to crystallize via random nucleation followed by progressive growth of crystals around nuclei (see figure \ref{fig:GuidedCrystallization} a). In this material, the change of phase from amorphous to crystalline produces an increase of the refractive index as well as a narrowing of the band gap, as displayed in figure \ref{fig:GuidedCrystallization} f. This leads to a clear change in the optical properties in the visible and near-infrared range, enabling us to easily monitor the formation and growth of crystals with a simple optical microscope. Monitoring crystals' formation and growth upon heating an Sb$_2$S$_3$ thin film at temperatures around the crystallization point confirms the randomness of nucleation sites apparition as shown in figure \ref{fig:GuidedCrystallization} a (see also supplementary video S1).

Beyond its inherent randomness, Sb$_2$S$_3$ crystallization process is also hindered by a limited number of nucleation sites, leading to a low crystal density per unit area, as shown in Fig. \ref{fig:GuidedCrystallization} a. A graphical estimate of the nucleation density (for these specific conditions of volume and temperature) yields a value of N$_d$ = 8.7x10$^{-5}$ µm$^{-2}$. Therefore, confining amorphous Sb$_2$S$_3$ thin films within micrometer-scale patterns of decreasing sizes -- achieved through photolithography and lift-off patterning -- gradually reduces the probability of finding a nucleation site. For square patterns whose sides are smaller than 10 $\mu$m, this probability drops to zero, leaving the material completely amorphous even after prolonged exposure to high temperatures (see Fig. \ref{fig:GuidedCrystallization} b) and supplementary information). This nucleation-limited crystallization presents a significant challenge for future integrated photonic devices, as many of the building blocks for tunable nanophotonics require nanostructured PCM-based elements, such as Sb$_2$S$_3$-based meta-atoms.

However, this dependence on the number of nucleation sites per unit volume can be transformed into an opportunity to achieve precise, on-demand control of crystallization on a chip. Our approach has conceptual similarities with the conventional bulk crystallization from the melt, where large-scale high-quality single crystals are grown with a specifically defined orientation, starting from an oriented germ \cite{raoux2008phase,uecker2014historical}. In our approach, we select a sufficiently large spatial region of the sample to ensure the appearance of several nucleation sites and use this area as a planar crystallization reservoir. This reservoir is then physically connected to channels that are used to guide the progressive planar crystallization to arbitrary locations. We capitalize on the isothermal time-crystallization method previously demonstrated \cite{taute2023programming} by maintaining a constant temperature near the crystallization threshold of Sb$_2$S$_3$. Under these conditions, the crystal gradually grows and propagates through the channels over time, as illustrated in figures \ref{fig:GuidedCrystallization} d and e. Since the crystal growth rate is both predictable and programmable for a given temperature, this guided crystallization strategy provides a simple yet robust means of controlling the PCM crystallization with high precision both in time and space.

\subsection{Multilevel programmable optical phase-shifter} \label{PIC}

\begin{figure*}[htbp!]
\centering\includegraphics[width=1\linewidth]{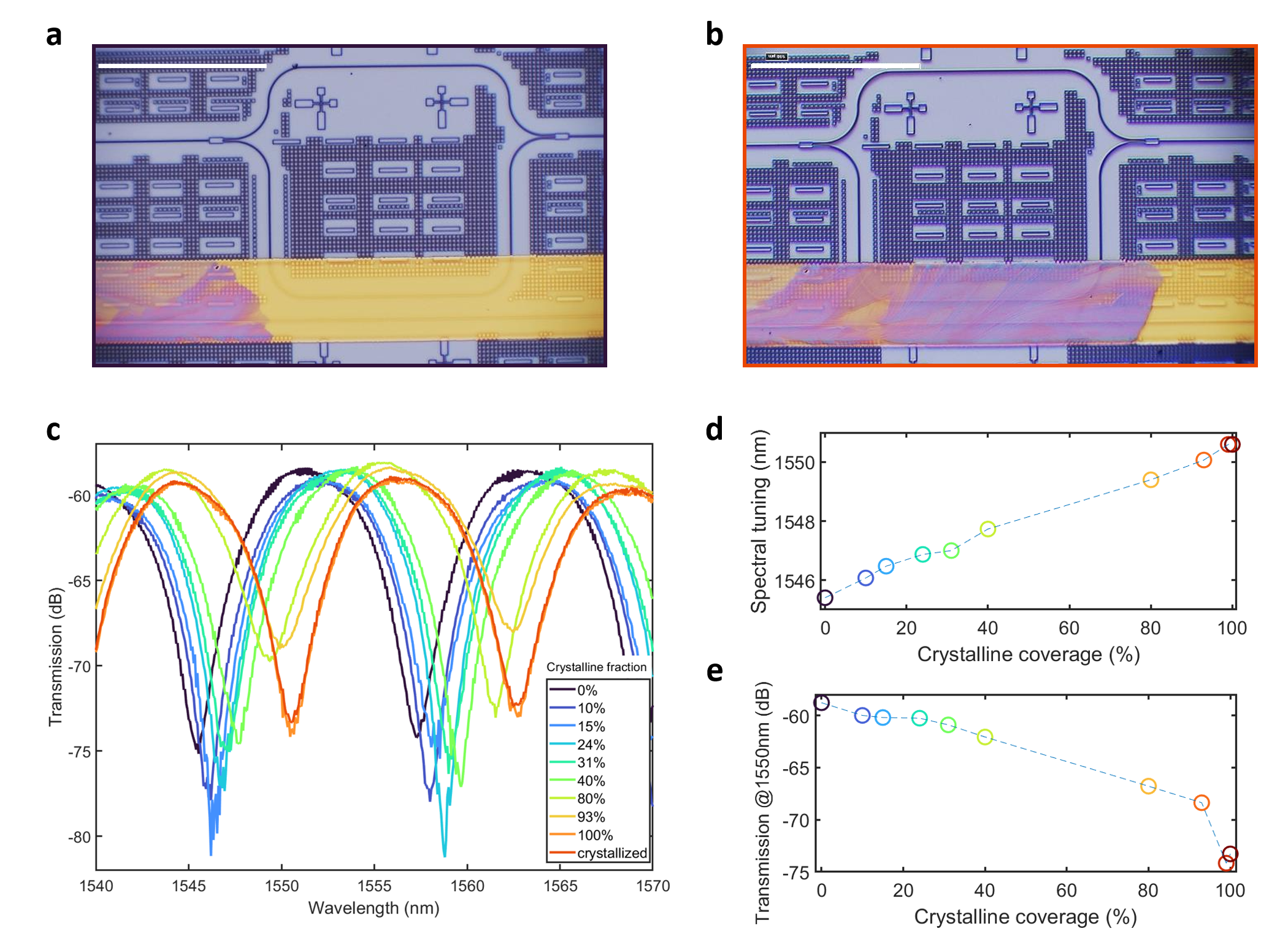}
\caption{\textbf{Experimental results on programmable MZI exploiting guided crystallization of Sb$_2$S$_3$.} a,b) Optical microscope view of the spatially-controlled planar crystal growth of Sb$_2$S$_3$, guided on top of one arm of the MZI. In these unprocessed images, the amorphous Sb$_2$S$_3$ appears yellow, while its crystalline phase materializes in magenta. The position of the crystal front is programmed via isothermal time-crystallization at 250°C. Scale bars: 100 µm. c) Raw output power of a single MZI for ten different programmed crystalline coverage of Sb$_2$S$_3$ (measured by graphical estimate from the microscope image) on the MZI arm. d) Spectral tuning and e) transmission modulation at 1550 nm, produced by the progressive crystallization of Sb$_2$S$_3$.}
\label{fig:MZI Results}
\end{figure*}

The first implementation of this concept in hybrid integrated photonic devices is demonstrated using a silicon-nitride-based Mach-Zehnder interferometer. A standard configuration for hybrid PCM-based waveguide devices involves integrating PCM patches on top of waveguides. As displayed in Fig. \ref{fig:GuidedCrystallization} d and \ref{fig:MZI Results} a, we employ a SiN waveguide platform, onto which we deposited a 50-nm-thick Sb$_2$S$_3$ layer (details can be found in the section Methods). 
Crystallizing the Sb$_2$S$_3$ patch significantly increases its refractive index (see figure \ref{fig:GuidedCrystallization} f, {where we display the refractive index n and extinction coefficient k of Sb$_2$S$_3$ for both the amorphous and crystalline phases}), leading to a change in the effective index of the hybrid waveguide mode (see Fig. S3 in the supplementary). In this configuration, the refractive index change between amorphous and crystalline phases modifies the effective index of the guided mode from n$^{amorph}_{eff}$= 1.608 to n$^{Crystal}_{eff}$ = 1.625. By precisely controlling the crystalline front of Sb$_2$S$_3$ on only one arm of the MZI, we can program the effective phase shift and, consequently, the phase difference at the output of the two guided modes forming the interferometer. 

As a first proof of principle, we exploit the guided crystallization within a simple linear Sb$_2$S$_3$ channel to program distinct non-volatile output states for the MZI. The key to reliably controlling and guiding the crystallization is to use a sufficiently narrow channel to have statistically one nucleus per channel. In our case, a 50 µm wide channel secures that condition (see supplementary Video S2). By heating the device to 250°C using a hot plate, after a few minutes of incubation, we create a nucleus that acts as a seed for the crystal growth, allowing the initially amorphous Sb$_2$S$_3$ to progressively crystallize along the channel until reaching the top of the MZI arm. We directly observe the crystal 'flow' within the guiding channel using an optical microscope, as the crystallization front is characterized by a magenta color seen in Fig. \ref{fig:MZI Results} a, resulting from structural color due to thin film interferences. In these specific conditions of volume and temperature, we measure a crystal growth speed of $\sim$ 16.5 µm/min (see the supplementary info S2 for more details). By simply stopping the heating at different times, we precisely control the position of the crystalline front of Sb$_2$S$_3$ relative to the MZI arm (see figure \ref{fig:MZI Results} a and b). At each step, we measure the optical transmission spectrum of the device in the near-infrared range (more details in the Methods section). As displayed in the transmission spectra in Fig. \ref{fig:MZI Results} c), this method enabled programming ten distinct levels in the output spectrum of the MZI. The overall modulation produced by the Sb$_2$S$_3$ crystallization results in a spectral tuning of the transmission peaks by 5 nm (Fig. \ref{fig:MZI Results} d) and a modulation of the transmission by 15 dB at 1550 nm (Fig. \ref{fig:MZI Results} e), and this, without specifically engineering the hybrid device to optimize its efficiency.

These states are non-volatile and were implemented using a simple hot plate. Once programmed, the device maintains its output state indefinitely unless reheated {to restart the crystal growth from where it stopped}. Note that only the phase of the transmitted mode is varying, while its amplitude remains largely unchanged, as the  Sb$_2$S$_3$ crystallization process induces minimal variation in the absorption coefficient. Importantly, the relatively moderate speed of crystal growth (16.5 µm/min, i.e. 275 nm/s) was purposefully chosen here to ease the programming stage and to precisely control the crystal front. Further optimizing the heating time and temperature would enable controlling the crystal front at the nanoscale and tailoring a continuum of intermediate levels between the extrema displayed in Fig. \ref{fig:MZI Results} c-e). Our tunable MZI results, therefore, highlight the potential of this approach for developing compact, non-volatile phase shifters with continuously programmable properties.

However, we also note important limitations of the proposed concept so far: i) we have no control on the location of the nucleus within the channel, and this one could very well appear on top of the device; ii) the crystal growth, even though directional and spatially-controlled, still results in polycrystalline layers, as can be seen from the different shades of colors in the crystals (see Fig. \ref{fig:MZI Results} a,b). In the following, we show how to circumvent these issues.

\begin{figure*}[htbp]
    \centering
    \includegraphics[width=1\textwidth]{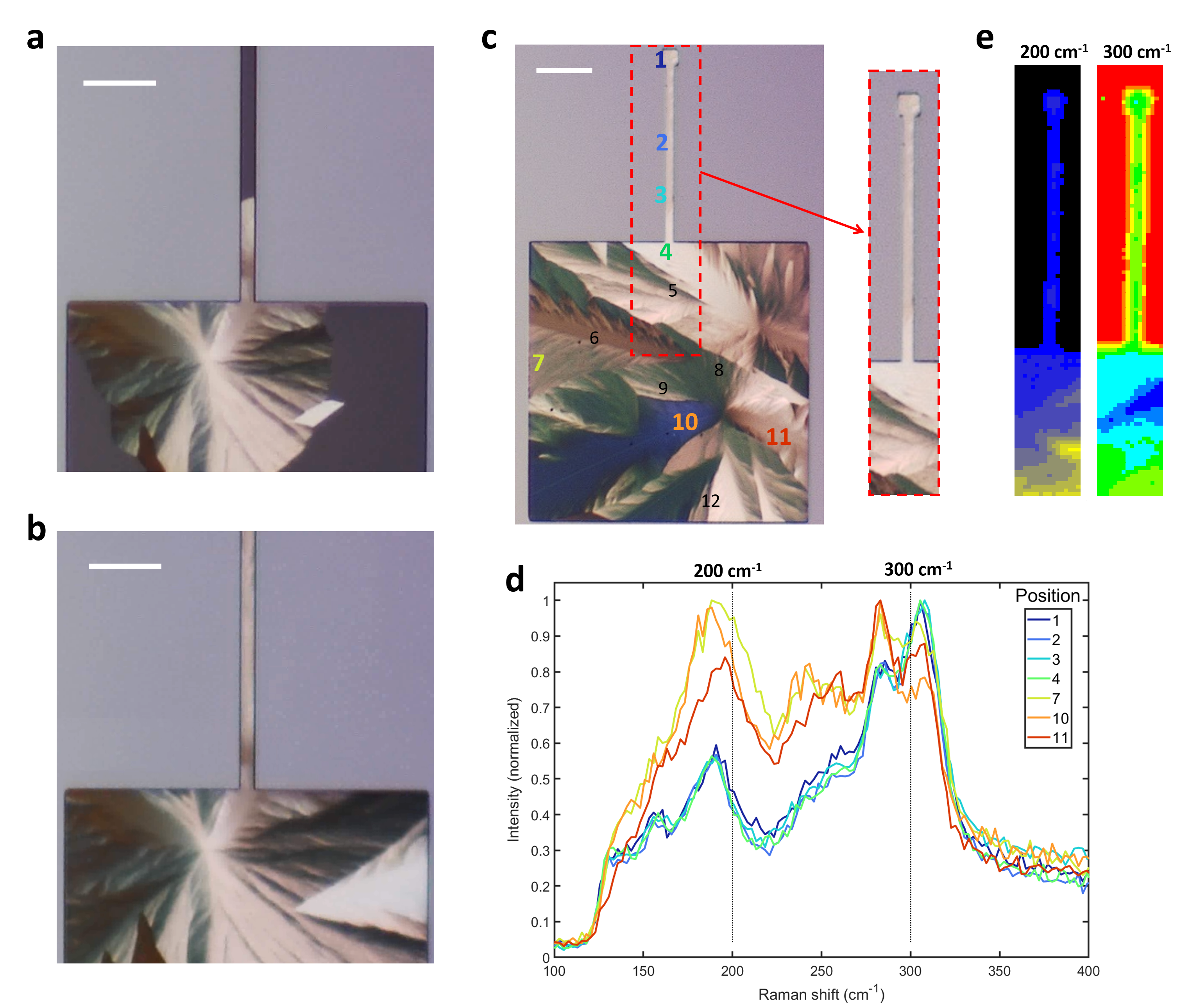}
    \caption{\textbf{Experimental demonstration of planar growth of optically homogeneous Sb$_2$S$_3$ crystals}:  (a,b) Polarized optical microscope images of the same patterned Sb$_2$S$_3$ device heated at 260°C for two different durations (scale bar: 20 $\mu$m). The device comprises a crystal reservoir, which is a large square of Sb$_2$S$_3$ (here 100 $\mu$m side) in which the probability for nucleation/crystallization is 100\%. This crystal reservoir is connected to a 2 $\mu$m-wide channel in which the probability for spontaneous crystallization is zero. The isothermal crystallization (constant temperature, 260°C) produces radial, uncontrolled growth of the crystal in the reservoir, while the same crystal is progressively guided in the channel as a function of time (see Video S3 for more details). c) Optical microscope image of another crystal reservoir connected to a 2µm wide channel and ending with a 5x5 µm pad. Such a small channel width enables filtering the crystal growth, yielding a presumably  homogeneous layer (zoomed-up image). (d) Each crystalline orientation of Sb$_2$S$_3$ presents a different Raman spectrum signature. This is particularly evident when measuring different zones of the reservoir (see e.g. zones 4, 7, and 10 and corresponding spectra). In the spectrum, the Raman shift around 200 cm$^{-1}$ strongly varies between the different regions and can therefore be used to map the crystalline domains (e) Micro-Raman mapping of the channel at Raman shifts of 200 cm$^{-1}$ and 300 cm$^{-1}$. We measure no variations of Raman intensity for both these wavenumbers along the channel and ending pad. This confirms the optically homogeneous nature of Sb$_2$S$_3$ in this region.}
    \label{fig:FilterCrystal}
\end{figure*}

\subsection{Filtering guided crystallization towards planar growth of optically homogeneous crystals}

The previous results demonstrate the ability to guide the crystallization of Sb$_2$S$_3$. However, as shown in the optical images in Fig. \ref{fig:MZI Results} a-b), this process results in a randomly oriented polycrystalline layer. Depending on the targeted type of optical components, polycrystallinity with different orientations might cause limitations to the device functionalities. Sb$_2$S$_3$ crystallizes in an orthorhombic structure, making it bi-anisotropic. As a result, a polycrystalline layer leads to an uncontrolled distribution of crystallographic orientations, causing random averaging of the associated refractive indices. Furthermore, grain boundaries between domains can act as scattering centers, introducing extrinsic optical losses. To address this issue, we further reduce the width of the guiding channel, aiming to filter out the guided crystallization process, towards limiting the in-plane growth to an optically homogeneous crystal. Furthermore, to deterministically choose the starting point of the crystal growth, we use crystal reservoir with dimensions sufficiently large to contain at least one crystal nucleus.

Figure \ref{fig:FilterCrystal} illustrates a 100 x 100 µm square crystal reservoir connected to a 2 µm-wide guiding channel, terminating in a 5x5 µm pad. Here again, the heat is maintained constant at 260°C, and we follow the isothermal crystallization in real-time. In figure \ref{fig:FilterCrystal} a-b), we show the crystal evolution at times t (a) and t + 1 minute (b). We see that the crystal grows spherically from the nucleus and propagates 'freely' in the reservoir, while on the contrary, its growth is guided in the channel (see also Supplementary Video S3). Note that the nucleation statistics are governed by the volume of the patterns (see Fig. S1 c in the supplement), hence the specific dimensions needed for the reservoir and channels should be optimized for each thickness of Sb$_2$S$_3$.

Using cross-polarization scheme in an optical microscope, we visualize different crystalline orientations, which, because of the optical anisotropy, appear as different colors. The polycrystalline nature of Sb$_2$S$_3$ within the reservoir is clearly seen in \ref{fig:FilterCrystal} a-c) via the different colors of the crystals. 
However, in both the channel and the pad, the measurement reveals only one color, suggesting the existence of an optically homogeneous crystal. To further confirm this, we employ Raman spectroscopy: due to its peculiar crystalline structure, Sb$_2$S$_3$ presents orientation-dependent Raman signatures, enabling point-by-point comparisons to assess whether a layer is poly- or mono-crystalline. Using a 405 nm laser, we performed several measurements at various locations on the devices presented in Figure \ref{fig:FilterCrystal} c (more details in the Methods section). The strongest Raman transitions are detected at 191, 283, and 301 cm$^{-1}$, corresponding to Ag vibrational modes, in agreement with previous studies \cite{liu2014first,gutierrez2022interlaboratory}. As explained above, the peak intensities of these transitions are highly dependent on crystal orientation. For example, within the crystal reservoir, we observed intensity variations of these Ag vibrational modes in the Raman spectra, confirming the presence of multiple orientations. However, in the channel and pad, the Raman spectra remain identical across measurements (see spectra marked 1, 2, 3 and 4 in Fig. \ref{fig:FilterCrystal} d), confirming that these regions contain well-defined, homogeneous crystalline orientations. 

For a more direct confirmation, we performed a Raman mapping scan over a 24-hour period for the specific region highlighted in Figure \ref{fig:FilterCrystal} c, focusing our measurements on two representative Ag Raman shifts: 200 and 300 cm$^{-1}$. For both wavenumbers, we measure no intensity variations along the channel and pad. 
This consistent spectral response confirms that the Sb$_2$S$_3$ layer in the channel is not polycrystalline. This result is further confirmed using polarized Raman measurements (see the supplement for more details).
While Raman mappings demonstrate the Sb$_2$S$_3$ channel is not polycrystalline, future works will help determine whether it is close to being monocrystalline or simply presenting a strong crystalline texture. In the following, we will define this oriented layer as 'optically homogeneous, a terminology we find intuitive to describe a crystal state whose effective optical properties resemble that of a monocrystal even though it is not fully monocrystalline \cite{barredo2014mechanical,fjellvåg2023pt,dai2008grain}.
The physics behind this optically homogeneous crystal growth is due to  the spatial restriction of the crystallization path, preventing the usual spherical growth around nuclei, and resulting in only a limited number of crystal growth direction allowed, along the channel. 

This in-plane guided filtered crystallization can be seen as a 2D planar version of existing conventional methods used to produce large-scale single crystals. For example, similarities can be found with the Czochralski method, in which a single crystal germ is used as a seed, introduced in a melted material, to form a large single crystal \cite{uecker2014historical}. 
Note that a recent work demonstrated on-chip Czochralski growth on unpatterned MoS$_2$ layers \cite{jiang2025two}. While the two processes bear similarities, our work extends on-chip Czochralski-like growth to another class of materials (chalcogenides), and goes beyond by providing temporal and spatial control on the change of phase in photonic devices. 

An even stronger analogy can be drawn between our proposed guided crystallization and the spiral grain selector in the so-called directional solidification method. In such a process, engineered 3D spirals are used to promote single crystal growth resulting from competitive grain growth during the solidification of a metal alloy \cite{dai2008grain,raza2019grain}. Nevertheless, our guided crystallization concept clearly stands out from these methods as it is based on a solid-to-solid transformation. This concept goes beyond the growth of crystalline material and leverages the controlled progressive transformation of the material itself (here Sb$_2$S$_3$) for spatially programming on-chip devices. 

As narrow guiding channels are needed to grow optically homogeneous crystals of Sb$_2$S$_3$, this method cannot be directly used to grow a wafer-scale monocrystal. However, we are not just limited to single straight narrow channels: we foresee that chip-scale devices could be programmed with optically homogeneous Sb$_2$S$_3$ crystals, as long as they comprise connected patterns smaller than the threshold area for nucleation. In the following, we show how this unlocks the potential of PCM for nanophotonic devices. 

\subsection{Non-volatile Programmable Optical Metasurfaces}

\begin{figure*}[htbp]
    \centering
    \includegraphics[width=1\textwidth]{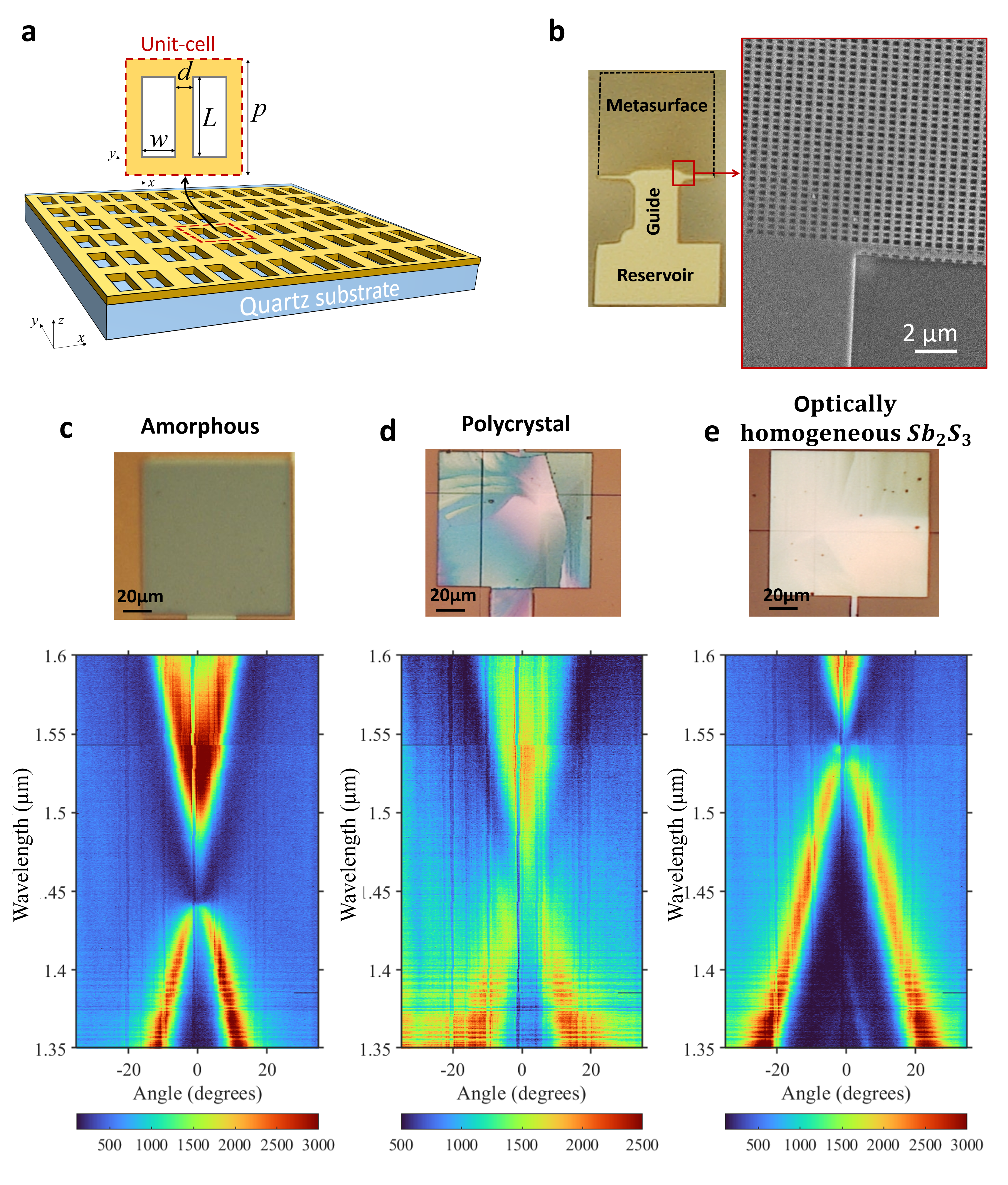}
    \caption{\textbf{Demonstration of Sb$_2$S$_3$-based metasurfaces with reconfigurable BIC exploiting guided growth of optically homogeneous crystals}:  (a) Perspective view of the designed metasurface with hole-like meta-atoms exhibiting an azimuthal symmetry rotation. In this design, $w$= 200 nm , $d$= 250 nm, $L$= 600 nm, $P$= 900 nm (b) Optical image of the system and close-up SEM view of the fabricated metasurface with the channel guiding the crystallization of the connected nanostructures. (c) Optical images and associated reflectivity bandstructure responses presenting a BIC dispersion when Sb$_2$S$_3$ is in its as-deposited amorphous state (c), polycrystalline state (d), and in an optically homogeneous crystal structure (e). The ‘quality’ of the crystal is governed by the width of the guiding channel: by restricting its width it acts as a ‘crystal filter’ limiting the number of crystals that seed the resulting growth inside the metasurface. We measure a large and clean spectral shift of the BIC only when the Sb$_2$S$_3$ layer is optically homogeneous}.
    \label{fig:BIC}
\end{figure*}

Building on our previous demonstrations in micron-scale channels, we apply our guided crystallization scheme to nanoscale patterns, addressing the key challenge of conceiving reconfigurable metasurfaces. Since isolated Sb$_2$S$_3$ patterns smaller than around 10$\mu$m rarely crystallize due to the limited availability of nucleation sites, this represents an issue to switch artificial meta-atoms, or structured materials, with typical dimensions on the order of a few hundred nanometers, generally used for the design of metasurfaces. To circumvent this fundamental limitation, we introduced an apertured metasurface design, where the meta-atoms are composed of nano-holes surrounded by Sb$_2$S$_3$. Using the same reservoir-assisted guided crystallization strategy, we achieve precise and controllable metasurface crystallization, unlocking innovative pathways for active and tunable photonics devices, down to the nanoscale.
Our metasurface design consists of a 2D periodic array of subwavelength rectangular nano-holes in a Sb$_2$S$_3$ film, arranged along the $x$ and $y$ directions, as shown in Figure \ref{fig:BIC} (a) (more details on the metasurface fabrication method and optical micro-reflectivity band diagram characterization are available in the Methods section). 

The experimental reflectivity bandstructure of the metasurface in the amorphous state is shown in Figure \ref{fig:BIC} (d). We observe the presence of photonic modes, including a bound state in the continuum (BIC) in the near-infrared range, associated with the vanishing transmission at k=0 and linewidth narrowing. 
The BIC mode is a non-radiative mode at normal incidence, exhibiting a quadratic decrease in the quality factor as a function of the increasing angle. This unique mode arises from the azimuthal symmetry rotation operator, which preserves the symmetry of the metasurface but does not allow direct excitation by normally-incident light. As a result, this mode remains decoupled from the far-field, resulting in a non-radiative resonance with an infinite Q-factor at normal incidence. Such peculiar high Q-factor resonances are highly desirable for various applications requiring strong light-matter interactions within a precisely engineered spectral region, e.g., for non-linear optics, bio-sensing, optical filtering, or integrated micro-lasers \cite{tseng2020dielectric,huang2020ultrafast,liu2019high,tian2023tunable}. As the spectral position of the BIC is governed by the material's refractive index, crystallizing the Sb$_2$S$_3$ metasurface from amorphous to crystalline should induce a spectral shift of the BIC. Enabling dynamic control over the spectral properties, tunable Sb$_2$S$_3$ BIC would significantly enhance the functionalities of reconfigurable photonic devices with non-volatile capabilities.

Figure \ref{fig:BIC} (c-e) top panels provide optical microscopic images comparing the visual aspect of a metasurface in its amorphous and crystallized states. The crystallization process begins in the reservoir, where a crystalline orientation is defined and initiated before propagating through the guided channel into the metasurface. As demonstrated previously, the channel width plays a crucial role in controlling the crystal growth in the metasurface. For sufficiently large channels, typically when their width exceeds 5 µm, we observed the propagation of multi-crystal flow entering the metasurface (Figure \ref{fig:BIC} d)). When the channel width size is reduced to 2 µm or less, a decreasing number of crystals can propagate and enter the metasurface, towards an optically homogeneous crystal in the metasurface (Figure \ref{fig:BIC} e)). 

Figures \ref{fig:BIC} c and e display the angular reflectivity distribution (see the Methods section for more details) for two different crystallization configurations: amorphous and optically homogeneous crystals. In both cases, a sharp BIC mode is observed in the band structure. Upon crystallization, the BIC undergoes a spectral redshift of $\Delta\lambda=100$nm, as shown in figure \ref{fig:BIC} (d,g). This spectral reconfigurability is reproduced by Rigorous Coupled Wave Analysis (RCWA) simulations (see Fig. S5 in the supplement), replicating the spectral shift by changing only the refractive index of  Sb$_2$S$_3$. Notably, this behavior is not observed in polycrystalline Sb$_2$S$_3$, as can be observed in Figure \ref{fig:BIC} d. In this measurement, we show that the BIC response with polycrystalline material is blurred, with only a small red shift of $\Delta\lambda=20$ nm compared to its amorphous counterpart. This demonstrates the critical importance of achieving an optically homogeneous crystal in the metasurface, as polycrystallinity introduces an averaging of refractive indices and disrupts the spatially extended phase coherence of BIC, diminishing its spectral shift.

The guided and filtered crystallization method overcomes two major challenges in the development of tunable Sb$_2$S$_3$ metasurfaces: i) it avoids the issue related to the scarcity of crystal grain formation, enabling crystallization of nanoscale patterns-- essential building blocks of metasurfaces; ii) Mimicking liquid-to-solid crystal growth process such as directional solidification or Czochralski method, it achieves optically homogeneous crystal growth. Preserving the optical coherence by preventing uncontrolled averaging of refractive indices in polycrystalline layers leads to high-quality reconfigurable and non-volatile functional photonic devices.

\section{Discussion}

Compared to previous works that have demonstrated reconfigurable photonic devices with Sb$_2$S$_3$ or Sb$_2$Se$_3$ using binary amorphous/crystalline modulation schemes \cite{delaney2021nonvolatile,fang2022ultra,yang2023non}, our spatially-controlled guided crystallization scheme introduces a breakthrough on the precise control over the crystalline state of PCMs at sub-micron precision level. To move towards multilevel tunability, two approaches were recently proposed: (i) laser-based 'pixel' by 'pixel' re-amorphization of PCM layer \cite{delaney2021nonvolatile}. While this method is elegant, it remains difficult to implement in fully industrialized systems. (ii) micro-heaters-based control of crystalline states of PCMs \cite{zhang2021electrically,yang2023non,chen2023non}: through a proper calibration of input electrical pulses, researchers demonstrated multilevel tuning of MZI. Given the intrinsic stochastic nature of crystal grain formation in Sb$_2$(S,Se)$_3$, there is no deterministic control over the position or propagation of the crystalline front in the devices. As a result, each device may exhibit a unique and unpredictable optical response, which might present scalability issues to ensure uniform performance across large-scale integrated photonic circuits presented in section \ref{PIC}.
Our experimental demonstration provides an alternative solution to predictably control the exact position of the crystal front within small patches of PCMs. 

One question related to this guided crystallization process is the speed limit of front propagation. In these initial demonstrations, we have purposefully chosen to make the transformation slow for a more precise programming of the devices (here $\sim$ 16.5 µm/min). This speed is clearly not a fundamental limit, and the crystallization kinetics can be directly controlled by the temperature. In a previous study, we have shown that the crystal growth of Sb$_2$S$_3$ follows the Avrami law and the speed of crystallization therefore exponentially increases with temperature \cite{taute2023programming}. In cases where a faster crystallization is needed, this parameter could easily be enhanced by increasing the temperature, up to 300°C (threshold temperature after which the Sb$_2$S$_3$ may crystallize differently \cite{gutierrez2022interlaboratory}). Although we did not investigate the fundamental speed limit, our previous study indicates a crystal growth rate ten times faster by increasing the temperature from 250°C to 270°C. Having this degree of freedom to control the speed of transformation is an important point: depending on the targeted photonic applications, precision can be a more critical factor than speed. For instance, achieving highly accurate, deterministic programming of multiple devices on a chip in programmable photonic circuits is essential to optimize global operation and fidelity. This is exactly similar to field-programmable gate arrays (FPGA), where the chip is configured before the system deployment. In the specific context of this work, a microsecond-scale reconfiguration time was perfectly acceptable, whereas an imprecise, non-optimized reconfiguration would have compromised the system's performance. 

Looking ahead, this progressive spatially-controlled planar crystal growth opens up exciting new research directions. To further reduce the footprint of the reservoir and enhance the local control of the crystallization, methods to deterministically create nuclei may be introduced, such as laser-induced heating or local doping to promote the nucleation and crystallization at specific locations of devices. Future development could integrate micro-heaters with strategically placed crystallization reservoirs to enable on-chip, programmable crystallization domains. Achieving controlled amorphization would further expand the reconfigurability of PCM-based photonic devices. This could readily be realized through rapid heating and quenching via integrated micro-heaters, allowing a full amorphization-reset process followed by recrystallization. Not to mention that the overall speed could be significantly improved by designing efficient heaters to confine the heat at specific locations, hence enabling rapid cycles. Our approach could thus unlock the ultimate goal of reversible, multi-level phase modulation.

\section{Conclusion}
In summary, we have proposed and demonstrated the novel concept of spatially-controlled planar crystal growth in low-loss PCMs, enabling the precise programming of integrated photonic devices and significantly enhancing the quality of reconfigurable nanophotonic devices. We have demonstrated the applicability of guided crystallization on the emerging PCM Sb$_2$S$_3$ in two different devices: a programmable optical phase shifter and a reconfigurable metasurface. By patterning an Sb$_2$S$_3$ patch onto one arm of a Mach-Zehnder interferometer, we successfully programmed ten distinct phase shifts, leveraging the refractive index change of the PCM during crystallization, all while maintaining negligible optical absorption. Going beyond conventional phase-change modulation, we experimentally demonstrated the original concept of in-plane guided growth of optically homogeneous Sb$_2$S$_3$ crystals, analogous to a planar and spatially-controlled version of conventional single crystal production processes such as grain selection in directional solidification or the Czochralski growth. The ability to precisely control the phase transition of Sb$_2$S$_3$ between amorphous and optically homogeneous states was leveraged in reconfigurable metasurfaces, enabling a $100~nm$ spectral modulation of a BIC mode, while preserving the metasurface functionalities, including optical coherence across a large area of the device. These results establish a new platform for high precision and efficient programmable/reconfigurable nanophotonic devices.

\section*{Methods}

\subsection*{Fabrication of hybrid MZI}
The silicon nitride-based waveguides were fabricated in an industrial foundry by STMicroelectronics. A 600-nm-thick silicon nitride (SiN) was deposited by physical vapor deposition and was subsequently patterned by deep-UV lithography and etched by reactive ion etching. SiN waveguides are 600 nm wide to ensure mono-mode operation at 1550nm. 

Next, a first 50 nm thick film of SiO$_2$ is deposited by electron beam evaporation on top of the base sample. We then patterned a thin film of 50 nm of Sb$_2$S$_3$ deposited by electron beam evaporation on top of one of the arms of the MZI using a lift-off process. The sample was subsequently capped using a 50 nm film of SiO$_2$ to avoid degradation of the film at high temperatures. 

\subsection*{Optical characterization of the MZI}
A lensed optical fiber was used to inject light via edge coupling from a broadband SLED light source into the MZI. Output light was then collected with another lensed fiber, and the resulting spectra were measured in an Optical Spectrum Analzyer (OSA). This allows for the measurement of the transmission spectrum of the MZI between 1200 nm and 1700 nm. The MZI spectrum was first measured at room temperature, the Sb$_2$S$_3$ film being in its as-deposited state. The sample was subsequently gradually crystallized by placing it on a temperature-controlled stage at the selected temperature. The sample was then allowed to cool down to room temperature before acquiring the MZI spectrum. This ensures that no crystallization occurs during the measurement and lifts the need to take thermo-optic effects into account.

\subsection*{Raman characterization}
Raman spectroscopy analysis was performed on an unpolarized Renishaw Invia Qontor system with 1.5 cm$^{-1}$ resolution and x20 objective at 405 nm wavelength (0,3mW/µm² 10s).  
Polarized Raman spectra were obtained using a 475 nm narrow linewidth pump laser, focused to sub-micron spot size using a 100x objective. The polarization at both the excitation and the detection channels were set using conventional polarizers.

\subsection*{Numerical simulations of Bound State in the Continuum metasurfaces}
Numerical simulations of BIC metasurfaces were conducted to analyze the reflectivity bandstructure of a metasurface based on Sb$_2$S$_3$. The strategy involved designing a metasurface with in-plane inversion symmetry using periodic subwavelength rectangle unit cell to achieve BiC resonances \cite{mikheeva2019photosensitive,koshelev2018asymmetric}. The working wavelengths window were performed by scaling the meta-atoms sizes and periodicity for both amorphous and crystalline state. The calculations were performed using rigorous coupled wave analysis (RCWA). The substrate was assumed to be a lossless quartz material with a fixed refractive index of 1.46. The metasurface structure was configured with a periodic boundary condition and excited by transverse electric (TE) incidence parallel to the long axis of the double-rectangles structure, utilizing air and Sb$_2$S$_3$ material contrast. The optical properties were considered, including the real and imaginary refractive indices of Sb$_2$S$_3$. To display the reflectivity spectra as a function of the wavevector, the angle of incidence was varied from -20° to 20°. The bandstructure response was designed for the telecom range, with the metasurface parameters set to a thickness of 155 nm, a period of 900 nm, a length of 600 nm, and a width of 200 nm.

\subsection*{Metasurface Nanofabrication}
The metasurfaces were fabricated by depositing 150nm amorphous Sb$_2$S$_3$ layer on a glass substrate using an electron beam evaporator at a pressure below 5.10$^{-6}$ mbar and a deposition rate below 0.7 A/sec. 
Then, we deposited 15nm SiN as a protective layer of the PCM using PECVD at a low temperature of 180°C using SiH$_4$ and N$_2$ gases. We then proceed with the e-beam lithography, using a double layer of Poly(methyl methacrylate) (PMMA) to allow enough thickness to transfer the patterns to the Sb$_2$S$_3$/ SiN layers. We spin-coated the PMMA at 4000 rpm for 30 seconds, baked it at 180°C for 1 minute and 30 seconds, and repeated the process to obtain 240nm thickness. The sample is then exposed using Electron Beam Lithography, following our designs. The resist is then developed using MIBK:IPA (1:1) for 1 min, followed by stopper IPA for 1 min. We transferred the pattern to the SiN and the Sb$_2$S$_3$ using ICP-RIE CHF$_3$:Ar:N$_2$ gas mixture and removed the residual resist using O$_2$ gas for 20s. Finally, we capped the device with 15nm SiN using PECVD.

\subsection*{Optical characterization of the metasurface}
Optical characterizations have been realized using a broadband light source (900nm-1.6µm) through free space, which is focused at normal incidence onto the sample using a 50x objective. The reflected light is Fourier transformed using a lens placed at the back focal plane of the objective and focused at the entrance slit of a spectrograph \cite{cueff2024fourier}. The light collected by the detector allows for angle-resolved spectroscopy, which is then normalized by the background and light reference using glass and mirror substrate for transmission and reflection measurement, respectively.

\section*{Funding}
We acknowledge funding from the French National Research Agency (ANR) under the  MetaOnDemand (ANR-20-CE24-0013).

\medskip

\bigskip

%%%%%%%%%%%%%%%%%%%%%%% References %%%%%%%%%%%%%%%%%%%%%%%%%

%%%%%%%%%% If using BibTeX:
\bibliography{Biblio_GuidedCrystallization}

\begin{thebibliography}{60}%
\makeatletter
\providecommand \@ifxundefined [1]{%
 \@ifx{#1\undefined}
}%
\providecommand \@ifnum [1]{%
 \ifnum #1\expandafter \@firstoftwo
 \else \expandafter \@secondoftwo
 \fi
}%
\providecommand \@ifx [1]{%
 \ifx #1\expandafter \@firstoftwo
 \else \expandafter \@secondoftwo
 \fi
}%
\providecommand \natexlab [1]{#1}%
\providecommand \enquote  [1]{``#1''}%
\providecommand \bibnamefont  [1]{#1}%
\providecommand \bibfnamefont [1]{#1}%
\providecommand \citenamefont [1]{#1}%
\providecommand \href@noop [0]{\@secondoftwo}%
\providecommand \href [0]{\begingroup \@sanitize@url \@href}%
\providecommand \@href[1]{\@@startlink{#1}\@@href}%
\providecommand \@@href[1]{\endgroup#1\@@endlink}%
\providecommand \@sanitize@url [0]{\catcode `\\12\catcode `\$12\catcode `\&12\catcode `\#12\catcode `\^12\catcode `\_12\catcode `\%12\relax}%
\providecommand \@@startlink[1]{}%
\providecommand \@@endlink[0]{}%
\providecommand \url  [0]{\begingroup\@sanitize@url \@url }%
\providecommand \@url [1]{\endgroup\@href {#1}{\urlprefix }}%
\providecommand \urlprefix  [0]{URL }%
\providecommand \Eprint [0]{\href }%
\providecommand \doibase [0]{https://doi.org/}%
\providecommand \selectlanguage [0]{\@gobble}%
\providecommand \bibinfo  [0]{\@secondoftwo}%
\providecommand \bibfield  [0]{\@secondoftwo}%
\providecommand \translation [1]{[#1]}%
\providecommand \BibitemOpen [0]{}%
\providecommand \bibitemStop [0]{}%
\providecommand \bibitemNoStop [0]{.\EOS\space}%
\providecommand \EOS [0]{\spacefactor3000\relax}%
\providecommand \BibitemShut  [1]{\csname bibitem#1\endcsname}%
\let\auto@bib@innerbib\@empty
%</preamble>
\bibitem [{\citenamefont {Genevet}\ \emph {et~al.}(2017)\citenamefont {Genevet}, \citenamefont {Capasso}, \citenamefont {Aieta}, \citenamefont {Khorasaninejad},\ and\ \citenamefont {Devlin}}]{Genevet_2015}%
  \BibitemOpen
  \bibfield  {author} {\bibinfo {author} {\bibfnamefont {P.}~\bibnamefont {Genevet}}, \bibinfo {author} {\bibfnamefont {F.}~\bibnamefont {Capasso}}, \bibinfo {author} {\bibfnamefont {F.}~\bibnamefont {Aieta}}, \bibinfo {author} {\bibfnamefont {M.}~\bibnamefont {Khorasaninejad}},\ and\ \bibinfo {author} {\bibfnamefont {R.}~\bibnamefont {Devlin}},\ }\bibfield  {title} {\bibinfo {title} {Recent advances in planar optics: from plasmonic to dielectric metasurfaces},\ }\href {https://doi.org/10.1364/OPTICA.4.000139} {\bibfield  {journal} {\bibinfo  {journal} {Optica}\ }\textbf {\bibinfo {volume} {4}},\ \bibinfo {pages} {139} (\bibinfo {year} {2017})}\BibitemShut {NoStop}%
\bibitem [{\citenamefont {Yu}\ and\ \citenamefont {Capasso}(2014)}]{yu2014flat}%
  \BibitemOpen
  \bibfield  {author} {\bibinfo {author} {\bibfnamefont {N.}~\bibnamefont {Yu}}\ and\ \bibinfo {author} {\bibfnamefont {F.}~\bibnamefont {Capasso}},\ }\bibfield  {title} {\bibinfo {title} {Flat optics with designer metasurfaces},\ }\href@noop {} {\bibfield  {journal} {\bibinfo  {journal} {Nature materials}\ }\textbf {\bibinfo {volume} {13}},\ \bibinfo {pages} {139} (\bibinfo {year} {2014})}\BibitemShut {NoStop}%
\bibitem [{\citenamefont {Shastri}\ \emph {et~al.}(2021)\citenamefont {Shastri}, \citenamefont {Tait}, \citenamefont {Ferreira~de Lima}, \citenamefont {Pernice}, \citenamefont {Bhaskaran}, \citenamefont {Wright},\ and\ \citenamefont {Prucnal}}]{shastri2021photonics}%
  \BibitemOpen
  \bibfield  {author} {\bibinfo {author} {\bibfnamefont {B.~J.}\ \bibnamefont {Shastri}}, \bibinfo {author} {\bibfnamefont {A.~N.}\ \bibnamefont {Tait}}, \bibinfo {author} {\bibfnamefont {T.}~\bibnamefont {Ferreira~de Lima}}, \bibinfo {author} {\bibfnamefont {W.~H.}\ \bibnamefont {Pernice}}, \bibinfo {author} {\bibfnamefont {H.}~\bibnamefont {Bhaskaran}}, \bibinfo {author} {\bibfnamefont {C.~D.}\ \bibnamefont {Wright}},\ and\ \bibinfo {author} {\bibfnamefont {P.~R.}\ \bibnamefont {Prucnal}},\ }\bibfield  {title} {\bibinfo {title} {Photonics for artificial intelligence and neuromorphic computing},\ }\href@noop {} {\bibfield  {journal} {\bibinfo  {journal} {Nature Photonics}\ }\textbf {\bibinfo {volume} {15}},\ \bibinfo {pages} {102} (\bibinfo {year} {2021})}\BibitemShut {NoStop}%
\bibitem [{\citenamefont {Wang}\ \emph {et~al.}(2024)\citenamefont {Wang}, \citenamefont {Hao}, \citenamefont {He}, \citenamefont {Xie}, \citenamefont {Liu}, \citenamefont {Tan}, \citenamefont {Li}, \citenamefont {Wang}, \citenamefont {Genevet}, \citenamefont {Luo} \emph {et~al.}}]{wang2024advances}%
  \BibitemOpen
  \bibfield  {author} {\bibinfo {author} {\bibfnamefont {X.}~\bibnamefont {Wang}}, \bibinfo {author} {\bibfnamefont {H.}~\bibnamefont {Hao}}, \bibinfo {author} {\bibfnamefont {X.}~\bibnamefont {He}}, \bibinfo {author} {\bibfnamefont {P.}~\bibnamefont {Xie}}, \bibinfo {author} {\bibfnamefont {J.}~\bibnamefont {Liu}}, \bibinfo {author} {\bibfnamefont {J.}~\bibnamefont {Tan}}, \bibinfo {author} {\bibfnamefont {H.}~\bibnamefont {Li}}, \bibinfo {author} {\bibfnamefont {H.}~\bibnamefont {Wang}}, \bibinfo {author} {\bibfnamefont {P.}~\bibnamefont {Genevet}}, \bibinfo {author} {\bibfnamefont {Y.}~\bibnamefont {Luo}}, \emph {et~al.},\ }\bibfield  {title} {\bibinfo {title} {Advances in information processing and biological imaging using flat optics},\ }\href@noop {} {\bibfield  {journal} {\bibinfo  {journal} {Nature Reviews Electrical Engineering}\ ,\ \bibinfo {pages} {1}} (\bibinfo {year} {2024})}\BibitemShut {NoStop}%
\bibitem [{\citenamefont {Cheben}\ \emph {et~al.}(2018)\citenamefont {Cheben}, \citenamefont {Halir}, \citenamefont {Schmid}, \citenamefont {Atwater},\ and\ \citenamefont {Smith}}]{cheben2018subwavelength}%
  \BibitemOpen
  \bibfield  {author} {\bibinfo {author} {\bibfnamefont {P.}~\bibnamefont {Cheben}}, \bibinfo {author} {\bibfnamefont {R.}~\bibnamefont {Halir}}, \bibinfo {author} {\bibfnamefont {J.~H.}\ \bibnamefont {Schmid}}, \bibinfo {author} {\bibfnamefont {H.~A.}\ \bibnamefont {Atwater}},\ and\ \bibinfo {author} {\bibfnamefont {D.~R.}\ \bibnamefont {Smith}},\ }\bibfield  {title} {\bibinfo {title} {Subwavelength integrated photonics},\ }\href@noop {} {\bibfield  {journal} {\bibinfo  {journal} {Nature}\ }\textbf {\bibinfo {volume} {560}},\ \bibinfo {pages} {565} (\bibinfo {year} {2018})}\BibitemShut {NoStop}%
\bibitem [{\citenamefont {Xie}\ \emph {et~al.}(2020)\citenamefont {Xie}, \citenamefont {Wang}, \citenamefont {Kan}, \citenamefont {Briere}, \citenamefont {Chen}, \citenamefont {Zhao}, \citenamefont {Delga}, \citenamefont {Ren}, \citenamefont {Chen}, \citenamefont {Xu},\ and\ \citenamefont {Genevet}}]{VCSEL_2020}%
  \BibitemOpen
  \bibfield  {author} {\bibinfo {author} {\bibfnamefont {Y.-Y.}\ \bibnamefont {Xie}}, \bibinfo {author} {\bibfnamefont {Q.-H.}\ \bibnamefont {Wang}}, \bibinfo {author} {\bibfnamefont {Q.}~\bibnamefont {Kan}}, \bibinfo {author} {\bibfnamefont {G.}~\bibnamefont {Briere}}, \bibinfo {author} {\bibfnamefont {P.-P.}\ \bibnamefont {Chen}}, \bibinfo {author} {\bibfnamefont {Z.-z.}\ \bibnamefont {Zhao}}, \bibinfo {author} {\bibfnamefont {A.}~\bibnamefont {Delga}}, \bibinfo {author} {\bibfnamefont {H.~R.}\ \bibnamefont {Ren}}, \bibinfo {author} {\bibfnamefont {H.-D.}\ \bibnamefont {Chen}}, \bibinfo {author} {\bibfnamefont {c.}~\bibnamefont {Xu}},\ and\ \bibinfo {author} {\bibfnamefont {P.}~\bibnamefont {Genevet}},\ }\bibfield  {title} {\bibinfo {title} {Metasurface-integrated vertical cavity surface-emitting lasers for programmable directional lasing emissions},\ }\href@noop {} {\bibfield  {journal} {\bibinfo  {journal} {NAture Nanotechnology}\ }\textbf {\bibinfo {volume} {15}},\ \bibinfo {pages} {125} (\bibinfo {year}
  {2020})}\BibitemShut {NoStop}%
\bibitem [{\citenamefont {Mikheeva}\ \emph {et~al.}(2022)\citenamefont {Mikheeva}, \citenamefont {Kyrou}, \citenamefont {Bentata}, \citenamefont {Khadir}, \citenamefont {Cueff},\ and\ \citenamefont {Genevet}}]{mikheeva2022space}%
  \BibitemOpen
  \bibfield  {author} {\bibinfo {author} {\bibfnamefont {E.}~\bibnamefont {Mikheeva}}, \bibinfo {author} {\bibfnamefont {C.}~\bibnamefont {Kyrou}}, \bibinfo {author} {\bibfnamefont {F.}~\bibnamefont {Bentata}}, \bibinfo {author} {\bibfnamefont {S.}~\bibnamefont {Khadir}}, \bibinfo {author} {\bibfnamefont {S.}~\bibnamefont {Cueff}},\ and\ \bibinfo {author} {\bibfnamefont {P.}~\bibnamefont {Genevet}},\ }\bibfield  {title} {\bibinfo {title} {Space and time modulations of light with metasurfaces: Recent progress and future prospects},\ }\href@noop {} {\bibfield  {journal} {\bibinfo  {journal} {ACS Photonics}\ }\textbf {\bibinfo {volume} {9}},\ \bibinfo {pages} {1458} (\bibinfo {year} {2022})}\BibitemShut {NoStop}%
\bibitem [{\citenamefont {Kuznetsov}\ \emph {et~al.}(2024)\citenamefont {Kuznetsov}, \citenamefont {Brongersma}, \citenamefont {Yao}, \citenamefont {Chen}, \citenamefont {Levy}, \citenamefont {Tsai}, \citenamefont {Zheludev}, \citenamefont {Faraon}, \citenamefont {Arbabi}, \citenamefont {Yu} \emph {et~al.}}]{kuznetsov2024roadmap}%
  \BibitemOpen
  \bibfield  {author} {\bibinfo {author} {\bibfnamefont {A.~I.}\ \bibnamefont {Kuznetsov}}, \bibinfo {author} {\bibfnamefont {M.~L.}\ \bibnamefont {Brongersma}}, \bibinfo {author} {\bibfnamefont {J.}~\bibnamefont {Yao}}, \bibinfo {author} {\bibfnamefont {M.~K.}\ \bibnamefont {Chen}}, \bibinfo {author} {\bibfnamefont {U.}~\bibnamefont {Levy}}, \bibinfo {author} {\bibfnamefont {D.~P.}\ \bibnamefont {Tsai}}, \bibinfo {author} {\bibfnamefont {N.~I.}\ \bibnamefont {Zheludev}}, \bibinfo {author} {\bibfnamefont {A.}~\bibnamefont {Faraon}}, \bibinfo {author} {\bibfnamefont {A.}~\bibnamefont {Arbabi}}, \bibinfo {author} {\bibfnamefont {N.}~\bibnamefont {Yu}}, \emph {et~al.},\ }\bibfield  {title} {\bibinfo {title} {Roadmap for optical metasurfaces},\ }\href@noop {} {\bibfield  {journal} {\bibinfo  {journal} {ACS photonics}\ }\textbf {\bibinfo {volume} {11}},\ \bibinfo {pages} {816} (\bibinfo {year} {2024})}\BibitemShut {NoStop}%
\bibitem [{\citenamefont {Brongersma}\ \emph {et~al.}(2025)\citenamefont {Brongersma}, \citenamefont {Pala}, \citenamefont {Altug}, \citenamefont {Capasso}, \citenamefont {Chen}, \citenamefont {Majumdar},\ and\ \citenamefont {Atwater}}]{brongersma2025second}%
  \BibitemOpen
  \bibfield  {author} {\bibinfo {author} {\bibfnamefont {M.~L.}\ \bibnamefont {Brongersma}}, \bibinfo {author} {\bibfnamefont {R.~A.}\ \bibnamefont {Pala}}, \bibinfo {author} {\bibfnamefont {H.}~\bibnamefont {Altug}}, \bibinfo {author} {\bibfnamefont {F.}~\bibnamefont {Capasso}}, \bibinfo {author} {\bibfnamefont {W.~T.}\ \bibnamefont {Chen}}, \bibinfo {author} {\bibfnamefont {A.}~\bibnamefont {Majumdar}},\ and\ \bibinfo {author} {\bibfnamefont {H.~A.}\ \bibnamefont {Atwater}},\ }\bibfield  {title} {\bibinfo {title} {The second optical metasurface revolution: moving from science to technology},\ }\href@noop {} {\bibfield  {journal} {\bibinfo  {journal} {Nature Reviews Electrical Engineering}\ ,\ \bibinfo {pages} {1}} (\bibinfo {year} {2025})}\BibitemShut {NoStop}%
\bibitem [{\citenamefont {Bogaerts}\ \emph {et~al.}(2020)\citenamefont {Bogaerts}, \citenamefont {P{\'e}rez}, \citenamefont {Capmany}, \citenamefont {Miller}, \citenamefont {Poon}, \citenamefont {Englund}, \citenamefont {Morichetti},\ and\ \citenamefont {Melloni}}]{bogaerts2020programmable}%
  \BibitemOpen
  \bibfield  {author} {\bibinfo {author} {\bibfnamefont {W.}~\bibnamefont {Bogaerts}}, \bibinfo {author} {\bibfnamefont {D.}~\bibnamefont {P{\'e}rez}}, \bibinfo {author} {\bibfnamefont {J.}~\bibnamefont {Capmany}}, \bibinfo {author} {\bibfnamefont {D.~A.}\ \bibnamefont {Miller}}, \bibinfo {author} {\bibfnamefont {J.}~\bibnamefont {Poon}}, \bibinfo {author} {\bibfnamefont {D.}~\bibnamefont {Englund}}, \bibinfo {author} {\bibfnamefont {F.}~\bibnamefont {Morichetti}},\ and\ \bibinfo {author} {\bibfnamefont {A.}~\bibnamefont {Melloni}},\ }\bibfield  {title} {\bibinfo {title} {Programmable photonic circuits},\ }\href@noop {} {\bibfield  {journal} {\bibinfo  {journal} {Nature}\ }\textbf {\bibinfo {volume} {586}},\ \bibinfo {pages} {207} (\bibinfo {year} {2020})}\BibitemShut {NoStop}%
\bibitem [{\citenamefont {Liu}\ \emph {et~al.}(2016)\citenamefont {Liu}, \citenamefont {Li}, \citenamefont {Guzzon}, \citenamefont {Norberg}, \citenamefont {Parker}, \citenamefont {Lu}, \citenamefont {Coldren},\ and\ \citenamefont {Yao}}]{liu2016fully}%
  \BibitemOpen
  \bibfield  {author} {\bibinfo {author} {\bibfnamefont {W.}~\bibnamefont {Liu}}, \bibinfo {author} {\bibfnamefont {M.}~\bibnamefont {Li}}, \bibinfo {author} {\bibfnamefont {R.~S.}\ \bibnamefont {Guzzon}}, \bibinfo {author} {\bibfnamefont {E.~J.}\ \bibnamefont {Norberg}}, \bibinfo {author} {\bibfnamefont {J.~S.}\ \bibnamefont {Parker}}, \bibinfo {author} {\bibfnamefont {M.}~\bibnamefont {Lu}}, \bibinfo {author} {\bibfnamefont {L.~A.}\ \bibnamefont {Coldren}},\ and\ \bibinfo {author} {\bibfnamefont {J.}~\bibnamefont {Yao}},\ }\bibfield  {title} {\bibinfo {title} {A fully reconfigurable photonic integrated signal processor},\ }\href@noop {} {\bibfield  {journal} {\bibinfo  {journal} {Nature Photonics}\ }\textbf {\bibinfo {volume} {10}},\ \bibinfo {pages} {190} (\bibinfo {year} {2016})}\BibitemShut {NoStop}%
\bibitem [{\citenamefont {P{\'e}rez}\ \emph {et~al.}(2017)\citenamefont {P{\'e}rez}, \citenamefont {Gasulla}, \citenamefont {Crudgington}, \citenamefont {Thomson}, \citenamefont {Khokhar}, \citenamefont {Li}, \citenamefont {Cao}, \citenamefont {Mashanovich},\ and\ \citenamefont {Capmany}}]{perez2017multipurpose}%
  \BibitemOpen
  \bibfield  {author} {\bibinfo {author} {\bibfnamefont {D.}~\bibnamefont {P{\'e}rez}}, \bibinfo {author} {\bibfnamefont {I.}~\bibnamefont {Gasulla}}, \bibinfo {author} {\bibfnamefont {L.}~\bibnamefont {Crudgington}}, \bibinfo {author} {\bibfnamefont {D.~J.}\ \bibnamefont {Thomson}}, \bibinfo {author} {\bibfnamefont {A.~Z.}\ \bibnamefont {Khokhar}}, \bibinfo {author} {\bibfnamefont {K.}~\bibnamefont {Li}}, \bibinfo {author} {\bibfnamefont {W.}~\bibnamefont {Cao}}, \bibinfo {author} {\bibfnamefont {G.~Z.}\ \bibnamefont {Mashanovich}},\ and\ \bibinfo {author} {\bibfnamefont {J.}~\bibnamefont {Capmany}},\ }\bibfield  {title} {\bibinfo {title} {Multipurpose silicon photonics signal processor core},\ }\href@noop {} {\bibfield  {journal} {\bibinfo  {journal} {Nature communications}\ }\textbf {\bibinfo {volume} {8}},\ \bibinfo {pages} {1} (\bibinfo {year} {2017})}\BibitemShut {NoStop}%
\bibitem [{\citenamefont {Li}\ \emph {et~al.}(2019)\citenamefont {Li}, \citenamefont {Xu}, \citenamefont {Maruthiyodan~Veetil}, \citenamefont {Valuckas}, \citenamefont {Paniagua-Dom{\'\i}nguez},\ and\ \citenamefont {Kuznetsov}}]{li2019phase}%
  \BibitemOpen
  \bibfield  {author} {\bibinfo {author} {\bibfnamefont {S.-Q.}\ \bibnamefont {Li}}, \bibinfo {author} {\bibfnamefont {X.}~\bibnamefont {Xu}}, \bibinfo {author} {\bibfnamefont {R.}~\bibnamefont {Maruthiyodan~Veetil}}, \bibinfo {author} {\bibfnamefont {V.}~\bibnamefont {Valuckas}}, \bibinfo {author} {\bibfnamefont {R.}~\bibnamefont {Paniagua-Dom{\'\i}nguez}},\ and\ \bibinfo {author} {\bibfnamefont {A.~I.}\ \bibnamefont {Kuznetsov}},\ }\bibfield  {title} {\bibinfo {title} {Phase-only transmissive spatial light modulator based on tunable dielectric metasurface},\ }\href@noop {} {\bibfield  {journal} {\bibinfo  {journal} {Science}\ }\textbf {\bibinfo {volume} {364}},\ \bibinfo {pages} {1087} (\bibinfo {year} {2019})}\BibitemShut {NoStop}%
\bibitem [{\citenamefont {Van~Iseghem}\ \emph {et~al.}(2022)\citenamefont {Van~Iseghem}, \citenamefont {Picavet}, \citenamefont {Takabayashi}, \citenamefont {Edinger}, \citenamefont {Khan}, \citenamefont {Verheyen}, \citenamefont {Quack}, \citenamefont {Gylfason}, \citenamefont {De~Buysser}, \citenamefont {Beeckman} \emph {et~al.}}]{van2022low}%
  \BibitemOpen
  \bibfield  {author} {\bibinfo {author} {\bibfnamefont {L.}~\bibnamefont {Van~Iseghem}}, \bibinfo {author} {\bibfnamefont {E.}~\bibnamefont {Picavet}}, \bibinfo {author} {\bibfnamefont {A.~Y.}\ \bibnamefont {Takabayashi}}, \bibinfo {author} {\bibfnamefont {P.}~\bibnamefont {Edinger}}, \bibinfo {author} {\bibfnamefont {U.}~\bibnamefont {Khan}}, \bibinfo {author} {\bibfnamefont {P.}~\bibnamefont {Verheyen}}, \bibinfo {author} {\bibfnamefont {N.}~\bibnamefont {Quack}}, \bibinfo {author} {\bibfnamefont {K.~B.}\ \bibnamefont {Gylfason}}, \bibinfo {author} {\bibfnamefont {K.}~\bibnamefont {De~Buysser}}, \bibinfo {author} {\bibfnamefont {J.}~\bibnamefont {Beeckman}}, \emph {et~al.},\ }\bibfield  {title} {\bibinfo {title} {Low power optical phase shifter using liquid crystal actuation on a silicon photonics platform},\ }\href@noop {} {\bibfield  {journal} {\bibinfo  {journal} {Optical Materials Express}\ }\textbf {\bibinfo {volume} {12}},\ \bibinfo {pages} {2181} (\bibinfo {year} {2022})}\BibitemShut {NoStop}%
\bibitem [{\citenamefont {Feng}\ \emph {et~al.}(2020)\citenamefont {Feng}, \citenamefont {Thomson}, \citenamefont {Mashanovich},\ and\ \citenamefont {Yan}}]{feng2020performance}%
  \BibitemOpen
  \bibfield  {author} {\bibinfo {author} {\bibfnamefont {Y.}~\bibnamefont {Feng}}, \bibinfo {author} {\bibfnamefont {D.~J.}\ \bibnamefont {Thomson}}, \bibinfo {author} {\bibfnamefont {G.~Z.}\ \bibnamefont {Mashanovich}},\ and\ \bibinfo {author} {\bibfnamefont {J.}~\bibnamefont {Yan}},\ }\bibfield  {title} {\bibinfo {title} {Performance analysis of a silicon noems device applied as an optical modulator based on a slot waveguide},\ }\href@noop {} {\bibfield  {journal} {\bibinfo  {journal} {Optics Express}\ }\textbf {\bibinfo {volume} {28}},\ \bibinfo {pages} {38206} (\bibinfo {year} {2020})}\BibitemShut {NoStop}%
\bibitem [{\citenamefont {Wuttig}\ \emph {et~al.}(2017)\citenamefont {Wuttig}, \citenamefont {Bhaskaran},\ and\ \citenamefont {Taubner}}]{wuttig2017phase}%
  \BibitemOpen
  \bibfield  {author} {\bibinfo {author} {\bibfnamefont {M.}~\bibnamefont {Wuttig}}, \bibinfo {author} {\bibfnamefont {H.}~\bibnamefont {Bhaskaran}},\ and\ \bibinfo {author} {\bibfnamefont {T.}~\bibnamefont {Taubner}},\ }\bibfield  {title} {\bibinfo {title} {Phase-change materials for non-volatile photonic applications},\ }\href@noop {} {\bibfield  {journal} {\bibinfo  {journal} {Nature photonics}\ }\textbf {\bibinfo {volume} {11}},\ \bibinfo {pages} {465} (\bibinfo {year} {2017})}\BibitemShut {NoStop}%
\bibitem [{\citenamefont {Abdollahramezani}\ \emph {et~al.}(2020)\citenamefont {Abdollahramezani}, \citenamefont {Hemmatyar}, \citenamefont {Taghinejad}, \citenamefont {Krasnok}, \citenamefont {Kiarashinejad}, \citenamefont {Zandehshahvar}, \citenamefont {Al{\`u}},\ and\ \citenamefont {Adibi}}]{abdollahramezani2020tunable}%
  \BibitemOpen
  \bibfield  {author} {\bibinfo {author} {\bibfnamefont {S.}~\bibnamefont {Abdollahramezani}}, \bibinfo {author} {\bibfnamefont {O.}~\bibnamefont {Hemmatyar}}, \bibinfo {author} {\bibfnamefont {H.}~\bibnamefont {Taghinejad}}, \bibinfo {author} {\bibfnamefont {A.}~\bibnamefont {Krasnok}}, \bibinfo {author} {\bibfnamefont {Y.}~\bibnamefont {Kiarashinejad}}, \bibinfo {author} {\bibfnamefont {M.}~\bibnamefont {Zandehshahvar}}, \bibinfo {author} {\bibfnamefont {A.}~\bibnamefont {Al{\`u}}},\ and\ \bibinfo {author} {\bibfnamefont {A.}~\bibnamefont {Adibi}},\ }\bibfield  {title} {\bibinfo {title} {Tunable nanophotonics enabled by chalcogenide phase-change materials},\ }\href@noop {} {\bibfield  {journal} {\bibinfo  {journal} {Nanophotonics}\ }\textbf {\bibinfo {volume} {9}},\ \bibinfo {pages} {1189} (\bibinfo {year} {2020})}\BibitemShut {NoStop}%
\bibitem [{\citenamefont {Raoux}\ \emph {et~al.}(2008)\citenamefont {Raoux}, \citenamefont {Burr}, \citenamefont {Breitwisch}, \citenamefont {Rettner}, \citenamefont {Chen}, \citenamefont {Shelby}, \citenamefont {Salinga}, \citenamefont {Krebs}, \citenamefont {Chen}, \citenamefont {Lung} \emph {et~al.}}]{raoux2008phase}%
  \BibitemOpen
  \bibfield  {author} {\bibinfo {author} {\bibfnamefont {S.}~\bibnamefont {Raoux}}, \bibinfo {author} {\bibfnamefont {G.~W.}\ \bibnamefont {Burr}}, \bibinfo {author} {\bibfnamefont {M.~J.}\ \bibnamefont {Breitwisch}}, \bibinfo {author} {\bibfnamefont {C.~T.}\ \bibnamefont {Rettner}}, \bibinfo {author} {\bibfnamefont {Y.-C.}\ \bibnamefont {Chen}}, \bibinfo {author} {\bibfnamefont {R.~M.}\ \bibnamefont {Shelby}}, \bibinfo {author} {\bibfnamefont {M.}~\bibnamefont {Salinga}}, \bibinfo {author} {\bibfnamefont {D.}~\bibnamefont {Krebs}}, \bibinfo {author} {\bibfnamefont {S.-H.}\ \bibnamefont {Chen}}, \bibinfo {author} {\bibfnamefont {H.-L.}\ \bibnamefont {Lung}}, \emph {et~al.},\ }\bibfield  {title} {\bibinfo {title} {Phase-change random access memory: A scalable technology},\ }\href@noop {} {\bibfield  {journal} {\bibinfo  {journal} {IBM Journal of Research and Development}\ }\textbf {\bibinfo {volume} {52}},\ \bibinfo {pages} {465} (\bibinfo {year} {2008})}\BibitemShut {NoStop}%
\bibitem [{\citenamefont {Gholipour}\ \emph {et~al.}(2013)\citenamefont {Gholipour}, \citenamefont {Zhang}, \citenamefont {MacDonald}, \citenamefont {Hewak},\ and\ \citenamefont {Zheludev}}]{gholipour2013all}%
  \BibitemOpen
  \bibfield  {author} {\bibinfo {author} {\bibfnamefont {B.}~\bibnamefont {Gholipour}}, \bibinfo {author} {\bibfnamefont {J.}~\bibnamefont {Zhang}}, \bibinfo {author} {\bibfnamefont {K.~F.}\ \bibnamefont {MacDonald}}, \bibinfo {author} {\bibfnamefont {D.~W.}\ \bibnamefont {Hewak}},\ and\ \bibinfo {author} {\bibfnamefont {N.~I.}\ \bibnamefont {Zheludev}},\ }\bibfield  {title} {\bibinfo {title} {An all-optical, non-volatile, bidirectional, phase-change meta-switch},\ }\href@noop {} {\bibfield  {journal} {\bibinfo  {journal} {Advanced materials}\ }\textbf {\bibinfo {volume} {25}},\ \bibinfo {pages} {3050} (\bibinfo {year} {2013})}\BibitemShut {NoStop}%
\bibitem [{\citenamefont {R{\'\i}os}\ \emph {et~al.}(2015)\citenamefont {R{\'\i}os}, \citenamefont {Stegmaier}, \citenamefont {Hosseini}, \citenamefont {Wang}, \citenamefont {Scherer}, \citenamefont {Wright}, \citenamefont {Bhaskaran},\ and\ \citenamefont {Pernice}}]{rios2015integrated}%
  \BibitemOpen
  \bibfield  {author} {\bibinfo {author} {\bibfnamefont {C.}~\bibnamefont {R{\'\i}os}}, \bibinfo {author} {\bibfnamefont {M.}~\bibnamefont {Stegmaier}}, \bibinfo {author} {\bibfnamefont {P.}~\bibnamefont {Hosseini}}, \bibinfo {author} {\bibfnamefont {D.}~\bibnamefont {Wang}}, \bibinfo {author} {\bibfnamefont {T.}~\bibnamefont {Scherer}}, \bibinfo {author} {\bibfnamefont {C.~D.}\ \bibnamefont {Wright}}, \bibinfo {author} {\bibfnamefont {H.}~\bibnamefont {Bhaskaran}},\ and\ \bibinfo {author} {\bibfnamefont {W.~H.}\ \bibnamefont {Pernice}},\ }\bibfield  {title} {\bibinfo {title} {Integrated all-photonic non-volatile multi-level memory},\ }\href@noop {} {\bibfield  {journal} {\bibinfo  {journal} {Nature photonics}\ }\textbf {\bibinfo {volume} {9}},\ \bibinfo {pages} {725} (\bibinfo {year} {2015})}\BibitemShut {NoStop}%
\bibitem [{\citenamefont {Cheng}\ \emph {et~al.}(2018)\citenamefont {Cheng}, \citenamefont {Rios}, \citenamefont {Youngblood}, \citenamefont {Wright}, \citenamefont {Pernice},\ and\ \citenamefont {Bhaskaran}}]{cheng2018device}%
  \BibitemOpen
  \bibfield  {author} {\bibinfo {author} {\bibfnamefont {Z.}~\bibnamefont {Cheng}}, \bibinfo {author} {\bibfnamefont {C.}~\bibnamefont {Rios}}, \bibinfo {author} {\bibfnamefont {N.}~\bibnamefont {Youngblood}}, \bibinfo {author} {\bibfnamefont {C.~D.}\ \bibnamefont {Wright}}, \bibinfo {author} {\bibfnamefont {W.~H.}\ \bibnamefont {Pernice}},\ and\ \bibinfo {author} {\bibfnamefont {H.}~\bibnamefont {Bhaskaran}},\ }\bibfield  {title} {\bibinfo {title} {Device-level photonic memories and logic applications using phase-change materials},\ }\href@noop {} {\bibfield  {journal} {\bibinfo  {journal} {Advanced Materials}\ }\textbf {\bibinfo {volume} {30}},\ \bibinfo {pages} {1802435} (\bibinfo {year} {2018})}\BibitemShut {NoStop}%
\bibitem [{\citenamefont {Howes}\ \emph {et~al.}(2020)\citenamefont {Howes}, \citenamefont {Zhu}, \citenamefont {Curie}, \citenamefont {Avila}, \citenamefont {Wheeler}, \citenamefont {Haglund},\ and\ \citenamefont {Valentine}}]{howes2020optical}%
  \BibitemOpen
  \bibfield  {author} {\bibinfo {author} {\bibfnamefont {A.}~\bibnamefont {Howes}}, \bibinfo {author} {\bibfnamefont {Z.}~\bibnamefont {Zhu}}, \bibinfo {author} {\bibfnamefont {D.}~\bibnamefont {Curie}}, \bibinfo {author} {\bibfnamefont {J.~R.}\ \bibnamefont {Avila}}, \bibinfo {author} {\bibfnamefont {V.~D.}\ \bibnamefont {Wheeler}}, \bibinfo {author} {\bibfnamefont {R.~F.}\ \bibnamefont {Haglund}},\ and\ \bibinfo {author} {\bibfnamefont {J.~G.}\ \bibnamefont {Valentine}},\ }\bibfield  {title} {\bibinfo {title} {Optical limiting based on huygens’ metasurfaces},\ }\href@noop {} {\bibfield  {journal} {\bibinfo  {journal} {Nano Letters}\ }\textbf {\bibinfo {volume} {20}},\ \bibinfo {pages} {4638} (\bibinfo {year} {2020})}\BibitemShut {NoStop}%
\bibitem [{\citenamefont {Tripathi}\ \emph {et~al.}(2021)\citenamefont {Tripathi}, \citenamefont {John}, \citenamefont {Kruk}, \citenamefont {Zhang}, \citenamefont {Nguyen}, \citenamefont {Berguiga}, \citenamefont {Romeo}, \citenamefont {Orobtchouk}, \citenamefont {Ramanathan}, \citenamefont {Kivshar} \emph {et~al.}}]{tripathi2021tunable}%
  \BibitemOpen
  \bibfield  {author} {\bibinfo {author} {\bibfnamefont {A.}~\bibnamefont {Tripathi}}, \bibinfo {author} {\bibfnamefont {J.}~\bibnamefont {John}}, \bibinfo {author} {\bibfnamefont {S.}~\bibnamefont {Kruk}}, \bibinfo {author} {\bibfnamefont {Z.}~\bibnamefont {Zhang}}, \bibinfo {author} {\bibfnamefont {H.~S.}\ \bibnamefont {Nguyen}}, \bibinfo {author} {\bibfnamefont {L.}~\bibnamefont {Berguiga}}, \bibinfo {author} {\bibfnamefont {P.~R.}\ \bibnamefont {Romeo}}, \bibinfo {author} {\bibfnamefont {R.}~\bibnamefont {Orobtchouk}}, \bibinfo {author} {\bibfnamefont {S.}~\bibnamefont {Ramanathan}}, \bibinfo {author} {\bibfnamefont {Y.}~\bibnamefont {Kivshar}}, \emph {et~al.},\ }\bibfield  {title} {\bibinfo {title} {Tunable mie-resonant dielectric metasurfaces based on vo2 phase-transition materials},\ }\href@noop {} {\bibfield  {journal} {\bibinfo  {journal} {ACS photonics}\ }\textbf {\bibinfo {volume} {8}},\ \bibinfo {pages} {1206} (\bibinfo {year} {2021})}\BibitemShut {NoStop}%
\bibitem [{\citenamefont {Cueff}\ \emph {et~al.}(2020)\citenamefont {Cueff}, \citenamefont {John}, \citenamefont {Zhang}, \citenamefont {Parra}, \citenamefont {Sun}, \citenamefont {Orobtchouk}, \citenamefont {Ramanathan},\ and\ \citenamefont {Sanchis}}]{cueff2020vo2}%
  \BibitemOpen
  \bibfield  {author} {\bibinfo {author} {\bibfnamefont {S.}~\bibnamefont {Cueff}}, \bibinfo {author} {\bibfnamefont {J.}~\bibnamefont {John}}, \bibinfo {author} {\bibfnamefont {Z.}~\bibnamefont {Zhang}}, \bibinfo {author} {\bibfnamefont {J.}~\bibnamefont {Parra}}, \bibinfo {author} {\bibfnamefont {J.}~\bibnamefont {Sun}}, \bibinfo {author} {\bibfnamefont {R.}~\bibnamefont {Orobtchouk}}, \bibinfo {author} {\bibfnamefont {S.}~\bibnamefont {Ramanathan}},\ and\ \bibinfo {author} {\bibfnamefont {P.}~\bibnamefont {Sanchis}},\ }\bibfield  {title} {\bibinfo {title} {Vo2 nanophotonics},\ }\href@noop {} {\bibfield  {journal} {\bibinfo  {journal} {APL Photonics}\ }\textbf {\bibinfo {volume} {5}},\ \bibinfo {pages} {110901} (\bibinfo {year} {2020})}\BibitemShut {NoStop}%
\bibitem [{\citenamefont {Cueff}\ \emph {et~al.}(2021)\citenamefont {Cueff}, \citenamefont {Taute}, \citenamefont {Bourgade}, \citenamefont {Lumeau}, \citenamefont {Monfray}, \citenamefont {Song}, \citenamefont {Genevet}, \citenamefont {Devif}, \citenamefont {Letartre},\ and\ \citenamefont {Berguiga}}]{cueff2021reconfigurable}%
  \BibitemOpen
  \bibfield  {author} {\bibinfo {author} {\bibfnamefont {S.}~\bibnamefont {Cueff}}, \bibinfo {author} {\bibfnamefont {A.}~\bibnamefont {Taute}}, \bibinfo {author} {\bibfnamefont {A.}~\bibnamefont {Bourgade}}, \bibinfo {author} {\bibfnamefont {J.}~\bibnamefont {Lumeau}}, \bibinfo {author} {\bibfnamefont {S.}~\bibnamefont {Monfray}}, \bibinfo {author} {\bibfnamefont {Q.}~\bibnamefont {Song}}, \bibinfo {author} {\bibfnamefont {P.}~\bibnamefont {Genevet}}, \bibinfo {author} {\bibfnamefont {B.}~\bibnamefont {Devif}}, \bibinfo {author} {\bibfnamefont {X.}~\bibnamefont {Letartre}},\ and\ \bibinfo {author} {\bibfnamefont {L.}~\bibnamefont {Berguiga}},\ }\bibfield  {title} {\bibinfo {title} {Reconfigurable flat optics with programmable reflection amplitude using lithography-free phase-change material ultra-thin films},\ }\href@noop {} {\bibfield  {journal} {\bibinfo  {journal} {Advanced Optical Materials}\ }\textbf {\bibinfo {volume} {9}},\ \bibinfo {pages} {2001291} (\bibinfo {year} {2021})}\BibitemShut {NoStop}%
\bibitem [{\citenamefont {Li}\ \emph {et~al.}(2016)\citenamefont {Li}, \citenamefont {Yang}, \citenamefont {Ma{\ss}}, \citenamefont {Hanss}, \citenamefont {Lewin}, \citenamefont {Michel}, \citenamefont {Wuttig},\ and\ \citenamefont {Taubner}}]{li2016reversible}%
  \BibitemOpen
  \bibfield  {author} {\bibinfo {author} {\bibfnamefont {P.}~\bibnamefont {Li}}, \bibinfo {author} {\bibfnamefont {X.}~\bibnamefont {Yang}}, \bibinfo {author} {\bibfnamefont {T.~W.}\ \bibnamefont {Ma{\ss}}}, \bibinfo {author} {\bibfnamefont {J.}~\bibnamefont {Hanss}}, \bibinfo {author} {\bibfnamefont {M.}~\bibnamefont {Lewin}}, \bibinfo {author} {\bibfnamefont {A.-K.~U.}\ \bibnamefont {Michel}}, \bibinfo {author} {\bibfnamefont {M.}~\bibnamefont {Wuttig}},\ and\ \bibinfo {author} {\bibfnamefont {T.}~\bibnamefont {Taubner}},\ }\bibfield  {title} {\bibinfo {title} {Reversible optical switching of highly confined phonon--polaritons with an ultrathin phase-change material},\ }\href@noop {} {\bibfield  {journal} {\bibinfo  {journal} {Nature materials}\ }\textbf {\bibinfo {volume} {15}},\ \bibinfo {pages} {870} (\bibinfo {year} {2016})}\BibitemShut {NoStop}%
\bibitem [{\citenamefont {Wang}\ \emph {et~al.}(2021)\citenamefont {Wang}, \citenamefont {Landreman}, \citenamefont {Schoen}, \citenamefont {Okabe}, \citenamefont {Marshall}, \citenamefont {Celano}, \citenamefont {Wong}, \citenamefont {Park},\ and\ \citenamefont {Brongersma}}]{wang2021electrical}%
  \BibitemOpen
  \bibfield  {author} {\bibinfo {author} {\bibfnamefont {Y.}~\bibnamefont {Wang}}, \bibinfo {author} {\bibfnamefont {P.}~\bibnamefont {Landreman}}, \bibinfo {author} {\bibfnamefont {D.}~\bibnamefont {Schoen}}, \bibinfo {author} {\bibfnamefont {K.}~\bibnamefont {Okabe}}, \bibinfo {author} {\bibfnamefont {A.}~\bibnamefont {Marshall}}, \bibinfo {author} {\bibfnamefont {U.}~\bibnamefont {Celano}}, \bibinfo {author} {\bibfnamefont {H.-S.~P.}\ \bibnamefont {Wong}}, \bibinfo {author} {\bibfnamefont {J.}~\bibnamefont {Park}},\ and\ \bibinfo {author} {\bibfnamefont {M.~L.}\ \bibnamefont {Brongersma}},\ }\bibfield  {title} {\bibinfo {title} {Electrical tuning of phase-change antennas and metasurfaces},\ }\href@noop {} {\bibfield  {journal} {\bibinfo  {journal} {Nature Nanotechnology}\ }\textbf {\bibinfo {volume} {16}},\ \bibinfo {pages} {667} (\bibinfo {year} {2021})}\BibitemShut {NoStop}%
\bibitem [{\citenamefont {Zhang}\ \emph {et~al.}(2025)\citenamefont {Zhang}, \citenamefont {Wang}, \citenamefont {Shen}, \citenamefont {Jana}, \citenamefont {Tan}, \citenamefont {Tian},\ and\ \citenamefont {Singh}}]{zhang2025chip}%
  \BibitemOpen
  \bibfield  {author} {\bibinfo {author} {\bibfnamefont {S.}~\bibnamefont {Zhang}}, \bibinfo {author} {\bibfnamefont {W.}~\bibnamefont {Wang}}, \bibinfo {author} {\bibfnamefont {Z.}~\bibnamefont {Shen}}, \bibinfo {author} {\bibfnamefont {S.}~\bibnamefont {Jana}}, \bibinfo {author} {\bibfnamefont {T.~C.}\ \bibnamefont {Tan}}, \bibinfo {author} {\bibfnamefont {Z.}~\bibnamefont {Tian}},\ and\ \bibinfo {author} {\bibfnamefont {R.}~\bibnamefont {Singh}},\ }\bibfield  {title} {\bibinfo {title} {On-chip non-volatile reconfigurable phase change topological photonics},\ }\href@noop {} {\bibfield  {journal} {\bibinfo  {journal} {Advanced Materials}\ }\textbf {\bibinfo {volume} {37}},\ \bibinfo {pages} {2418510} (\bibinfo {year} {2025})}\BibitemShut {NoStop}%
\bibitem [{\citenamefont {Zhang}\ \emph {et~al.}(2019)\citenamefont {Zhang}, \citenamefont {Chou}, \citenamefont {Li}, \citenamefont {Li}, \citenamefont {Du}, \citenamefont {Yadav}, \citenamefont {Zhou}, \citenamefont {Shalaginov}, \citenamefont {Fang}, \citenamefont {Zhong} \emph {et~al.}}]{zhang2019broadband}%
  \BibitemOpen
  \bibfield  {author} {\bibinfo {author} {\bibfnamefont {Y.}~\bibnamefont {Zhang}}, \bibinfo {author} {\bibfnamefont {J.~B.}\ \bibnamefont {Chou}}, \bibinfo {author} {\bibfnamefont {J.}~\bibnamefont {Li}}, \bibinfo {author} {\bibfnamefont {H.}~\bibnamefont {Li}}, \bibinfo {author} {\bibfnamefont {Q.}~\bibnamefont {Du}}, \bibinfo {author} {\bibfnamefont {A.}~\bibnamefont {Yadav}}, \bibinfo {author} {\bibfnamefont {S.}~\bibnamefont {Zhou}}, \bibinfo {author} {\bibfnamefont {M.~Y.}\ \bibnamefont {Shalaginov}}, \bibinfo {author} {\bibfnamefont {Z.}~\bibnamefont {Fang}}, \bibinfo {author} {\bibfnamefont {H.}~\bibnamefont {Zhong}}, \emph {et~al.},\ }\bibfield  {title} {\bibinfo {title} {Broadband transparent optical phase change materials for high-performance nonvolatile photonics},\ }\href@noop {} {\bibfield  {journal} {\bibinfo  {journal} {Nature communications}\ }\textbf {\bibinfo {volume} {10}},\ \bibinfo {pages} {1} (\bibinfo {year} {2019})}\BibitemShut {NoStop}%
\bibitem [{\citenamefont {Dong}\ \emph {et~al.}(2019)\citenamefont {Dong}, \citenamefont {Liu}, \citenamefont {Behera}, \citenamefont {Lu}, \citenamefont {Ng}, \citenamefont {Sreekanth}, \citenamefont {Zhou}, \citenamefont {Yang},\ and\ \citenamefont {Simpson}}]{dong2019wide}%
  \BibitemOpen
  \bibfield  {author} {\bibinfo {author} {\bibfnamefont {W.}~\bibnamefont {Dong}}, \bibinfo {author} {\bibfnamefont {H.}~\bibnamefont {Liu}}, \bibinfo {author} {\bibfnamefont {J.~K.}\ \bibnamefont {Behera}}, \bibinfo {author} {\bibfnamefont {L.}~\bibnamefont {Lu}}, \bibinfo {author} {\bibfnamefont {R.~J.}\ \bibnamefont {Ng}}, \bibinfo {author} {\bibfnamefont {K.~V.}\ \bibnamefont {Sreekanth}}, \bibinfo {author} {\bibfnamefont {X.}~\bibnamefont {Zhou}}, \bibinfo {author} {\bibfnamefont {J.~K.}\ \bibnamefont {Yang}},\ and\ \bibinfo {author} {\bibfnamefont {R.~E.}\ \bibnamefont {Simpson}},\ }\bibfield  {title} {\bibinfo {title} {Wide bandgap phase change material tuned visible photonics},\ }\href@noop {} {\bibfield  {journal} {\bibinfo  {journal} {Advanced Functional Materials}\ }\textbf {\bibinfo {volume} {29}},\ \bibinfo {pages} {1806181} (\bibinfo {year} {2019})}\BibitemShut {NoStop}%
\bibitem [{\citenamefont {Delaney}\ \emph {et~al.}(2020)\citenamefont {Delaney}, \citenamefont {Zeimpekis}, \citenamefont {Lawson}, \citenamefont {Hewak},\ and\ \citenamefont {Muskens}}]{delaney2020new}%
  \BibitemOpen
  \bibfield  {author} {\bibinfo {author} {\bibfnamefont {M.}~\bibnamefont {Delaney}}, \bibinfo {author} {\bibfnamefont {I.}~\bibnamefont {Zeimpekis}}, \bibinfo {author} {\bibfnamefont {D.}~\bibnamefont {Lawson}}, \bibinfo {author} {\bibfnamefont {D.~W.}\ \bibnamefont {Hewak}},\ and\ \bibinfo {author} {\bibfnamefont {O.~L.}\ \bibnamefont {Muskens}},\ }\bibfield  {title} {\bibinfo {title} {A new family of ultralow loss reversible phase-change materials for photonic integrated circuits: Sb2s3 and sb2se3},\ }\href@noop {} {\bibfield  {journal} {\bibinfo  {journal} {Advanced functional materials}\ }\textbf {\bibinfo {volume} {30}},\ \bibinfo {pages} {2002447} (\bibinfo {year} {2020})}\BibitemShut {NoStop}%
\bibitem [{\citenamefont {Biega{\'n}ski}\ \emph {et~al.}(2024)\citenamefont {Biega{\'n}ski}, \citenamefont {Perestjuk}, \citenamefont {Armand}, \citenamefont {Della~Torre}, \citenamefont {Laprais}, \citenamefont {Saint-Girons}, \citenamefont {Reboud}, \citenamefont {Hartmann}, \citenamefont {Tortai}, \citenamefont {Moreau} \emph {et~al.}}]{bieganski2024sb}%
  \BibitemOpen
  \bibfield  {author} {\bibinfo {author} {\bibfnamefont {A.}~\bibnamefont {Biega{\'n}ski}}, \bibinfo {author} {\bibfnamefont {M.}~\bibnamefont {Perestjuk}}, \bibinfo {author} {\bibfnamefont {R.}~\bibnamefont {Armand}}, \bibinfo {author} {\bibfnamefont {A.}~\bibnamefont {Della~Torre}}, \bibinfo {author} {\bibfnamefont {C.}~\bibnamefont {Laprais}}, \bibinfo {author} {\bibfnamefont {G.}~\bibnamefont {Saint-Girons}}, \bibinfo {author} {\bibfnamefont {V.}~\bibnamefont {Reboud}}, \bibinfo {author} {\bibfnamefont {J.-M.}\ \bibnamefont {Hartmann}}, \bibinfo {author} {\bibfnamefont {J.-H.}\ \bibnamefont {Tortai}}, \bibinfo {author} {\bibfnamefont {A.}~\bibnamefont {Moreau}}, \emph {et~al.},\ }\bibfield  {title} {\bibinfo {title} {Sb 2 s 3 as a low-loss phase-change material for mid-ir photonics},\ }\href@noop {} {\bibfield  {journal} {\bibinfo  {journal} {Optical Materials Express}\ }\textbf {\bibinfo {volume} {14}},\ \bibinfo {pages} {862} (\bibinfo {year} {2024})}\BibitemShut {NoStop}%
\bibitem [{\citenamefont {Fang}\ \emph {et~al.}(2021)\citenamefont {Fang}, \citenamefont {Zheng}, \citenamefont {Saxena}, \citenamefont {Whitehead}, \citenamefont {Chen},\ and\ \citenamefont {Majumdar}}]{fang2021non}%
  \BibitemOpen
  \bibfield  {author} {\bibinfo {author} {\bibfnamefont {Z.}~\bibnamefont {Fang}}, \bibinfo {author} {\bibfnamefont {J.}~\bibnamefont {Zheng}}, \bibinfo {author} {\bibfnamefont {A.}~\bibnamefont {Saxena}}, \bibinfo {author} {\bibfnamefont {J.}~\bibnamefont {Whitehead}}, \bibinfo {author} {\bibfnamefont {Y.}~\bibnamefont {Chen}},\ and\ \bibinfo {author} {\bibfnamefont {A.}~\bibnamefont {Majumdar}},\ }\bibfield  {title} {\bibinfo {title} {Non-volatile reconfigurable integrated photonics enabled by broadband low-loss phase change material},\ }\href@noop {} {\bibfield  {journal} {\bibinfo  {journal} {Advanced Optical Materials}\ }\textbf {\bibinfo {volume} {9}},\ \bibinfo {pages} {2002049} (\bibinfo {year} {2021})}\BibitemShut {NoStop}%
\bibitem [{\citenamefont {Chen}\ \emph {et~al.}(2023)\citenamefont {Chen}, \citenamefont {Fang}, \citenamefont {Perez}, \citenamefont {Miller}, \citenamefont {Kumari}, \citenamefont {Saxena}, \citenamefont {Zheng}, \citenamefont {Geiger}, \citenamefont {Goodson},\ and\ \citenamefont {Majumdar}}]{chen2023non}%
  \BibitemOpen
  \bibfield  {author} {\bibinfo {author} {\bibfnamefont {R.}~\bibnamefont {Chen}}, \bibinfo {author} {\bibfnamefont {Z.}~\bibnamefont {Fang}}, \bibinfo {author} {\bibfnamefont {C.}~\bibnamefont {Perez}}, \bibinfo {author} {\bibfnamefont {F.}~\bibnamefont {Miller}}, \bibinfo {author} {\bibfnamefont {K.}~\bibnamefont {Kumari}}, \bibinfo {author} {\bibfnamefont {A.}~\bibnamefont {Saxena}}, \bibinfo {author} {\bibfnamefont {J.}~\bibnamefont {Zheng}}, \bibinfo {author} {\bibfnamefont {S.~J.}\ \bibnamefont {Geiger}}, \bibinfo {author} {\bibfnamefont {K.~E.}\ \bibnamefont {Goodson}},\ and\ \bibinfo {author} {\bibfnamefont {A.}~\bibnamefont {Majumdar}},\ }\bibfield  {title} {\bibinfo {title} {Non-volatile electrically programmable integrated photonics with a 5-bit operation},\ }\href@noop {} {\bibfield  {journal} {\bibinfo  {journal} {Nature Communications}\ }\textbf {\bibinfo {volume} {14}},\ \bibinfo {pages} {3465} (\bibinfo {year} {2023})}\BibitemShut {NoStop}%
\bibitem [{\citenamefont {Fang}\ \emph {et~al.}(2024)\citenamefont {Fang}, \citenamefont {Chen}, \citenamefont {Froch}, \citenamefont {Tanguy}, \citenamefont {Khan}, \citenamefont {Wu}, \citenamefont {Tara}, \citenamefont {Manna}, \citenamefont {Sharp}, \citenamefont {Munley}, \citenamefont {Miller}, \citenamefont {Zhao}, \citenamefont {Geiger}, \citenamefont {Bohringer}, \citenamefont {Reynolds}, \citenamefont {Pop},\ and\ \citenamefont {Majumdar}}]{fang2024nonvolatile}%
  \BibitemOpen
  \bibfield  {author} {\bibinfo {author} {\bibfnamefont {Z.}~\bibnamefont {Fang}}, \bibinfo {author} {\bibfnamefont {R.}~\bibnamefont {Chen}}, \bibinfo {author} {\bibfnamefont {J.~E.}\ \bibnamefont {Froch}}, \bibinfo {author} {\bibfnamefont {Q.~A.}\ \bibnamefont {Tanguy}}, \bibinfo {author} {\bibfnamefont {A.~I.}\ \bibnamefont {Khan}}, \bibinfo {author} {\bibfnamefont {X.}~\bibnamefont {Wu}}, \bibinfo {author} {\bibfnamefont {V.}~\bibnamefont {Tara}}, \bibinfo {author} {\bibfnamefont {A.}~\bibnamefont {Manna}}, \bibinfo {author} {\bibfnamefont {D.}~\bibnamefont {Sharp}}, \bibinfo {author} {\bibfnamefont {C.}~\bibnamefont {Munley}}, \bibinfo {author} {\bibfnamefont {F.}~\bibnamefont {Miller}}, \bibinfo {author} {\bibfnamefont {Y.}~\bibnamefont {Zhao}}, \bibinfo {author} {\bibfnamefont {S.}~\bibnamefont {Geiger}}, \bibinfo {author} {\bibfnamefont {K.~F.}\ \bibnamefont {Bohringer}}, \bibinfo {author} {\bibfnamefont {M.~S.}\ \bibnamefont {Reynolds}}, \bibinfo {author} {\bibfnamefont {E.}~\bibnamefont {Pop}},\
  and\ \bibinfo {author} {\bibfnamefont {A.}~\bibnamefont {Majumdar}},\ }\bibfield  {title} {\bibinfo {title} {Nonvolatile phase-only transmissive spatial light modulator with electrical addressability of individual pixels},\ }\href@noop {} {\bibfield  {journal} {\bibinfo  {journal} {ACS nano}\ }\textbf {\bibinfo {volume} {18}},\ \bibinfo {pages} {11245} (\bibinfo {year} {2024})}\BibitemShut {NoStop}%
\bibitem [{\citenamefont {Moitra}\ \emph {et~al.}(2023)\citenamefont {Moitra}, \citenamefont {Wang}, \citenamefont {Liang}, \citenamefont {Lu}, \citenamefont {Poh}, \citenamefont {Mass}, \citenamefont {Simpson}, \citenamefont {Kuznetsov},\ and\ \citenamefont {Paniagua-Dominguez}}]{moitra2023programmable}%
  \BibitemOpen
  \bibfield  {author} {\bibinfo {author} {\bibfnamefont {P.}~\bibnamefont {Moitra}}, \bibinfo {author} {\bibfnamefont {Y.}~\bibnamefont {Wang}}, \bibinfo {author} {\bibfnamefont {X.}~\bibnamefont {Liang}}, \bibinfo {author} {\bibfnamefont {L.}~\bibnamefont {Lu}}, \bibinfo {author} {\bibfnamefont {A.}~\bibnamefont {Poh}}, \bibinfo {author} {\bibfnamefont {T.~W.}\ \bibnamefont {Mass}}, \bibinfo {author} {\bibfnamefont {R.~E.}\ \bibnamefont {Simpson}}, \bibinfo {author} {\bibfnamefont {A.~I.}\ \bibnamefont {Kuznetsov}},\ and\ \bibinfo {author} {\bibfnamefont {R.}~\bibnamefont {Paniagua-Dominguez}},\ }\bibfield  {title} {\bibinfo {title} {Programmable wavefront control in the visible spectrum using low-loss chalcogenide phase-change metasurfaces},\ }\href@noop {} {\bibfield  {journal} {\bibinfo  {journal} {Advanced Materials}\ }\textbf {\bibinfo {volume} {35}},\ \bibinfo {pages} {2205367} (\bibinfo {year} {2023})}\BibitemShut {NoStop}%
\bibitem [{\citenamefont {Delaney}\ \emph {et~al.}(2021)\citenamefont {Delaney}, \citenamefont {Zeimpekis}, \citenamefont {Du}, \citenamefont {Yan}, \citenamefont {Banakar}, \citenamefont {Thomson}, \citenamefont {Hewak},\ and\ \citenamefont {Muskens}}]{delaney2021nonvolatile}%
  \BibitemOpen
  \bibfield  {author} {\bibinfo {author} {\bibfnamefont {M.}~\bibnamefont {Delaney}}, \bibinfo {author} {\bibfnamefont {I.}~\bibnamefont {Zeimpekis}}, \bibinfo {author} {\bibfnamefont {H.}~\bibnamefont {Du}}, \bibinfo {author} {\bibfnamefont {X.}~\bibnamefont {Yan}}, \bibinfo {author} {\bibfnamefont {M.}~\bibnamefont {Banakar}}, \bibinfo {author} {\bibfnamefont {D.~J.}\ \bibnamefont {Thomson}}, \bibinfo {author} {\bibfnamefont {D.~W.}\ \bibnamefont {Hewak}},\ and\ \bibinfo {author} {\bibfnamefont {O.~L.}\ \bibnamefont {Muskens}},\ }\bibfield  {title} {\bibinfo {title} {Nonvolatile programmable silicon photonics using an ultralow-loss sb2se3 phase change material},\ }\href@noop {} {\bibfield  {journal} {\bibinfo  {journal} {Science Advances}\ }\textbf {\bibinfo {volume} {7}},\ \bibinfo {pages} {eabg3500} (\bibinfo {year} {2021})}\BibitemShut {NoStop}%
\bibitem [{\citenamefont {R{\'\i}os}\ \emph {et~al.}(2022)\citenamefont {R{\'\i}os}, \citenamefont {Du}, \citenamefont {Zhang}, \citenamefont {Popescu}, \citenamefont {Shalaginov}, \citenamefont {Miller}, \citenamefont {Roberts}, \citenamefont {Kang}, \citenamefont {Richardson}, \citenamefont {Gu} \emph {et~al.}}]{rios2022ultra}%
  \BibitemOpen
  \bibfield  {author} {\bibinfo {author} {\bibfnamefont {C.}~\bibnamefont {R{\'\i}os}}, \bibinfo {author} {\bibfnamefont {Q.}~\bibnamefont {Du}}, \bibinfo {author} {\bibfnamefont {Y.}~\bibnamefont {Zhang}}, \bibinfo {author} {\bibfnamefont {C.-C.}\ \bibnamefont {Popescu}}, \bibinfo {author} {\bibfnamefont {M.~Y.}\ \bibnamefont {Shalaginov}}, \bibinfo {author} {\bibfnamefont {P.}~\bibnamefont {Miller}}, \bibinfo {author} {\bibfnamefont {C.}~\bibnamefont {Roberts}}, \bibinfo {author} {\bibfnamefont {M.}~\bibnamefont {Kang}}, \bibinfo {author} {\bibfnamefont {K.~A.}\ \bibnamefont {Richardson}}, \bibinfo {author} {\bibfnamefont {T.}~\bibnamefont {Gu}}, \emph {et~al.},\ }\bibfield  {title} {\bibinfo {title} {Ultra-compact nonvolatile phase shifter based on electrically reprogrammable transparent phase change materials},\ }\href@noop {} {\bibfield  {journal} {\bibinfo  {journal} {PhotoniX}\ }\textbf {\bibinfo {volume} {3}},\ \bibinfo {pages} {26} (\bibinfo {year} {2022})}\BibitemShut {NoStop}%
\bibitem [{\citenamefont {Fang}\ \emph {et~al.}(2022)\citenamefont {Fang}, \citenamefont {Chen}, \citenamefont {Zheng}, \citenamefont {Khan}, \citenamefont {Neilson}, \citenamefont {Geiger}, \citenamefont {Callahan}, \citenamefont {Moebius}, \citenamefont {Saxena}, \citenamefont {Chen} \emph {et~al.}}]{fang2022ultra}%
  \BibitemOpen
  \bibfield  {author} {\bibinfo {author} {\bibfnamefont {Z.}~\bibnamefont {Fang}}, \bibinfo {author} {\bibfnamefont {R.}~\bibnamefont {Chen}}, \bibinfo {author} {\bibfnamefont {J.}~\bibnamefont {Zheng}}, \bibinfo {author} {\bibfnamefont {A.~I.}\ \bibnamefont {Khan}}, \bibinfo {author} {\bibfnamefont {K.~M.}\ \bibnamefont {Neilson}}, \bibinfo {author} {\bibfnamefont {S.~J.}\ \bibnamefont {Geiger}}, \bibinfo {author} {\bibfnamefont {D.~M.}\ \bibnamefont {Callahan}}, \bibinfo {author} {\bibfnamefont {M.~G.}\ \bibnamefont {Moebius}}, \bibinfo {author} {\bibfnamefont {A.}~\bibnamefont {Saxena}}, \bibinfo {author} {\bibfnamefont {M.~E.}\ \bibnamefont {Chen}}, \emph {et~al.},\ }\bibfield  {title} {\bibinfo {title} {Ultra-low-energy programmable non-volatile silicon photonics based on phase-change materials with graphene heaters},\ }\href@noop {} {\bibfield  {journal} {\bibinfo  {journal} {Nature nanotechnology}\ }\textbf {\bibinfo {volume} {17}},\ \bibinfo {pages} {842} (\bibinfo {year} {2022})}\BibitemShut
  {NoStop}%
\bibitem [{\citenamefont {Yang}\ \emph {et~al.}(2023)\citenamefont {Yang}, \citenamefont {Lu}, \citenamefont {Li}, \citenamefont {Wu}, \citenamefont {Li}, \citenamefont {Chen},\ and\ \citenamefont {Zhou}}]{yang2023non}%
  \BibitemOpen
  \bibfield  {author} {\bibinfo {author} {\bibfnamefont {X.}~\bibnamefont {Yang}}, \bibinfo {author} {\bibfnamefont {L.}~\bibnamefont {Lu}}, \bibinfo {author} {\bibfnamefont {Y.}~\bibnamefont {Li}}, \bibinfo {author} {\bibfnamefont {Y.}~\bibnamefont {Wu}}, \bibinfo {author} {\bibfnamefont {Z.}~\bibnamefont {Li}}, \bibinfo {author} {\bibfnamefont {J.}~\bibnamefont {Chen}},\ and\ \bibinfo {author} {\bibfnamefont {L.}~\bibnamefont {Zhou}},\ }\bibfield  {title} {\bibinfo {title} {Non-volatile optical switch element enabled by low-loss phase change material},\ }\href@noop {} {\bibfield  {journal} {\bibinfo  {journal} {Advanced Functional Materials}\ ,\ \bibinfo {pages} {2304601}} (\bibinfo {year} {2023})}\BibitemShut {NoStop}%
\bibitem [{\citenamefont {Tara}\ \emph {et~al.}(2024)\citenamefont {Tara}, \citenamefont {Chen}, \citenamefont {Froch}, \citenamefont {Fang}, \citenamefont {Fang}, \citenamefont {Audhkhasi}, \citenamefont {Choi},\ and\ \citenamefont {Majumdar}}]{tara2024non}%
  \BibitemOpen
  \bibfield  {author} {\bibinfo {author} {\bibfnamefont {V.}~\bibnamefont {Tara}}, \bibinfo {author} {\bibfnamefont {R.}~\bibnamefont {Chen}}, \bibinfo {author} {\bibfnamefont {J.~E.}\ \bibnamefont {Froch}}, \bibinfo {author} {\bibfnamefont {Z.}~\bibnamefont {Fang}}, \bibinfo {author} {\bibfnamefont {J.}~\bibnamefont {Fang}}, \bibinfo {author} {\bibfnamefont {R.}~\bibnamefont {Audhkhasi}}, \bibinfo {author} {\bibfnamefont {M.}~\bibnamefont {Choi}},\ and\ \bibinfo {author} {\bibfnamefont {A.}~\bibnamefont {Majumdar}},\ }\bibfield  {title} {\bibinfo {title} {Non-volatile reconfigurable transmissive notch filter using wide bandgap phase change material antimony sulfide},\ }\href@noop {} {\bibfield  {journal} {\bibinfo  {journal} {IEEE Journal of Selected Topics in Quantum Electronics}\ } (\bibinfo {year} {2024})}\BibitemShut {NoStop}%
\bibitem [{\citenamefont {Wu}\ \emph {et~al.}(2024)\citenamefont {Wu}, \citenamefont {Deng}, \citenamefont {Huang}, \citenamefont {Yu}, \citenamefont {Takeuchi}, \citenamefont {Rios~Ocampo},\ and\ \citenamefont {Li}}]{wu2024freeform}%
  \BibitemOpen
  \bibfield  {author} {\bibinfo {author} {\bibfnamefont {C.}~\bibnamefont {Wu}}, \bibinfo {author} {\bibfnamefont {H.}~\bibnamefont {Deng}}, \bibinfo {author} {\bibfnamefont {Y.-S.}\ \bibnamefont {Huang}}, \bibinfo {author} {\bibfnamefont {H.}~\bibnamefont {Yu}}, \bibinfo {author} {\bibfnamefont {I.}~\bibnamefont {Takeuchi}}, \bibinfo {author} {\bibfnamefont {C.~A.}\ \bibnamefont {Rios~Ocampo}},\ and\ \bibinfo {author} {\bibfnamefont {M.}~\bibnamefont {Li}},\ }\bibfield  {title} {\bibinfo {title} {Freeform direct-write and rewritable photonic integrated circuits in phase-change thin films},\ }\href@noop {} {\bibfield  {journal} {\bibinfo  {journal} {Science Advances}\ }\textbf {\bibinfo {volume} {10}},\ \bibinfo {pages} {eadk1361} (\bibinfo {year} {2024})}\BibitemShut {NoStop}%
\bibitem [{\citenamefont {Laprais}\ \emph {et~al.}(2024)\citenamefont {Laprais}, \citenamefont {Zrounba}, \citenamefont {Bouvier}, \citenamefont {Blanchard}, \citenamefont {Bugnet}, \citenamefont {Gassenq}, \citenamefont {Guti{\'e}rrez}, \citenamefont {Vazquez-Miranda}, \citenamefont {Espinoza}, \citenamefont {Thiesen} \emph {et~al.}}]{laprais2024reversible}%
  \BibitemOpen
  \bibfield  {author} {\bibinfo {author} {\bibfnamefont {C.}~\bibnamefont {Laprais}}, \bibinfo {author} {\bibfnamefont {C.}~\bibnamefont {Zrounba}}, \bibinfo {author} {\bibfnamefont {J.}~\bibnamefont {Bouvier}}, \bibinfo {author} {\bibfnamefont {N.}~\bibnamefont {Blanchard}}, \bibinfo {author} {\bibfnamefont {M.}~\bibnamefont {Bugnet}}, \bibinfo {author} {\bibfnamefont {A.}~\bibnamefont {Gassenq}}, \bibinfo {author} {\bibfnamefont {Y.}~\bibnamefont {Guti{\'e}rrez}}, \bibinfo {author} {\bibfnamefont {S.}~\bibnamefont {Vazquez-Miranda}}, \bibinfo {author} {\bibfnamefont {S.}~\bibnamefont {Espinoza}}, \bibinfo {author} {\bibfnamefont {P.}~\bibnamefont {Thiesen}}, \emph {et~al.},\ }\bibfield  {title} {\bibinfo {title} {Reversible single-pulse laser-induced phase change of sb2s3 thin films: Multi-physics modeling and experimental demonstrations},\ }\href@noop {} {\bibfield  {journal} {\bibinfo  {journal} {Advanced Optical Materials}\ }\textbf {\bibinfo {volume} {12}},\ \bibinfo {pages} {2401214} (\bibinfo {year}
  {2024})}\BibitemShut {NoStop}%
\bibitem [{\citenamefont {Taute}\ \emph {et~al.}(2023)\citenamefont {Taute}, \citenamefont {Al-Jibouri}, \citenamefont {Laprais}, \citenamefont {Monfray}, \citenamefont {Lumeau}, \citenamefont {Moreau}, \citenamefont {Letartre}, \citenamefont {Baboux}, \citenamefont {Saint-Girons}, \citenamefont {Berguiga} \emph {et~al.}}]{taute2023programming}%
  \BibitemOpen
  \bibfield  {author} {\bibinfo {author} {\bibfnamefont {A.}~\bibnamefont {Taute}}, \bibinfo {author} {\bibfnamefont {S.}~\bibnamefont {Al-Jibouri}}, \bibinfo {author} {\bibfnamefont {C.}~\bibnamefont {Laprais}}, \bibinfo {author} {\bibfnamefont {S.}~\bibnamefont {Monfray}}, \bibinfo {author} {\bibfnamefont {J.}~\bibnamefont {Lumeau}}, \bibinfo {author} {\bibfnamefont {A.}~\bibnamefont {Moreau}}, \bibinfo {author} {\bibfnamefont {X.}~\bibnamefont {Letartre}}, \bibinfo {author} {\bibfnamefont {N.}~\bibnamefont {Baboux}}, \bibinfo {author} {\bibfnamefont {G.}~\bibnamefont {Saint-Girons}}, \bibinfo {author} {\bibfnamefont {L.}~\bibnamefont {Berguiga}}, \emph {et~al.},\ }\bibfield  {title} {\bibinfo {title} {Programming multilevel crystallization states in phase-change-material thin films},\ }\href@noop {} {\bibfield  {journal} {\bibinfo  {journal} {Optical Materials Express}\ }\textbf {\bibinfo {volume} {13}},\ \bibinfo {pages} {3113} (\bibinfo {year} {2023})}\BibitemShut {NoStop}%
\bibitem [{\citenamefont {Dai}\ \emph {et~al.}(2008)\citenamefont {Dai}, \citenamefont {Gebelin}, \citenamefont {Newell}, \citenamefont {Reed}, \citenamefont {D’Souza}, \citenamefont {Brown},\ and\ \citenamefont {Dong}}]{dai2008grain}%
  \BibitemOpen
  \bibfield  {author} {\bibinfo {author} {\bibfnamefont {H.}~\bibnamefont {Dai}}, \bibinfo {author} {\bibfnamefont {J.}~\bibnamefont {Gebelin}}, \bibinfo {author} {\bibfnamefont {M.}~\bibnamefont {Newell}}, \bibinfo {author} {\bibfnamefont {R.}~\bibnamefont {Reed}}, \bibinfo {author} {\bibfnamefont {N.}~\bibnamefont {D’Souza}}, \bibinfo {author} {\bibfnamefont {P.}~\bibnamefont {Brown}},\ and\ \bibinfo {author} {\bibfnamefont {H.}~\bibnamefont {Dong}},\ }\bibfield  {title} {\bibinfo {title} {Grain selection during solidification in spiral grain selector},\ }\href@noop {} {\bibfield  {journal} {\bibinfo  {journal} {Superalloys}\ }\textbf {\bibinfo {volume} {2008}},\ \bibinfo {pages} {367} (\bibinfo {year} {2008})}\BibitemShut {NoStop}%
\bibitem [{\citenamefont {Raza}\ \emph {et~al.}(2019)\citenamefont {Raza}, \citenamefont {Wasim}, \citenamefont {Hussain}, \citenamefont {Sajid},\ and\ \citenamefont {Jahanzaib}}]{raza2019grain}%
  \BibitemOpen
  \bibfield  {author} {\bibinfo {author} {\bibfnamefont {M.~H.}\ \bibnamefont {Raza}}, \bibinfo {author} {\bibfnamefont {A.}~\bibnamefont {Wasim}}, \bibinfo {author} {\bibfnamefont {S.}~\bibnamefont {Hussain}}, \bibinfo {author} {\bibfnamefont {M.}~\bibnamefont {Sajid}},\ and\ \bibinfo {author} {\bibfnamefont {M.}~\bibnamefont {Jahanzaib}},\ }\bibfield  {title} {\bibinfo {title} {Grain selection and crystal orientation in single-crystal casting: State of the art},\ }\href@noop {} {\bibfield  {journal} {\bibinfo  {journal} {Crystal Research and Technology}\ }\textbf {\bibinfo {volume} {54}},\ \bibinfo {pages} {1800177} (\bibinfo {year} {2019})}\BibitemShut {NoStop}%
\bibitem [{\citenamefont {Uecker}(2014)}]{uecker2014historical}%
  \BibitemOpen
  \bibfield  {author} {\bibinfo {author} {\bibfnamefont {R.}~\bibnamefont {Uecker}},\ }\bibfield  {title} {\bibinfo {title} {The historical development of the czochralski method},\ }\href@noop {} {\bibfield  {journal} {\bibinfo  {journal} {Journal of Crystal Growth}\ }\textbf {\bibinfo {volume} {401}},\ \bibinfo {pages} {7} (\bibinfo {year} {2014})}\BibitemShut {NoStop}%
\bibitem [{\citenamefont {Liu}\ \emph {et~al.}(2014)\citenamefont {Liu}, \citenamefont {Chua}, \citenamefont {Sum},\ and\ \citenamefont {Gan}}]{liu2014first}%
  \BibitemOpen
  \bibfield  {author} {\bibinfo {author} {\bibfnamefont {Y.}~\bibnamefont {Liu}}, \bibinfo {author} {\bibfnamefont {K.~T.~E.}\ \bibnamefont {Chua}}, \bibinfo {author} {\bibfnamefont {T.~C.}\ \bibnamefont {Sum}},\ and\ \bibinfo {author} {\bibfnamefont {C.~K.}\ \bibnamefont {Gan}},\ }\bibfield  {title} {\bibinfo {title} {First-principles study of the lattice dynamics of sb 2 s 3},\ }\href@noop {} {\bibfield  {journal} {\bibinfo  {journal} {Physical Chemistry Chemical Physics}\ }\textbf {\bibinfo {volume} {16}},\ \bibinfo {pages} {345} (\bibinfo {year} {2014})}\BibitemShut {NoStop}%
\bibitem [{\citenamefont {Guti{\'e}rrez}\ \emph {et~al.}(2022)\citenamefont {Guti{\'e}rrez}, \citenamefont {Ovvyan}, \citenamefont {Santos}, \citenamefont {Juan}, \citenamefont {Rosales}, \citenamefont {Junquera}, \citenamefont {Garc{\'\i}a-Fern{\'a}ndez}, \citenamefont {Dicorato}, \citenamefont {Giangregorio}, \citenamefont {Dilonardo} \emph {et~al.}}]{gutierrez2022interlaboratory}%
  \BibitemOpen
  \bibfield  {author} {\bibinfo {author} {\bibfnamefont {Y.}~\bibnamefont {Guti{\'e}rrez}}, \bibinfo {author} {\bibfnamefont {A.~P.}\ \bibnamefont {Ovvyan}}, \bibinfo {author} {\bibfnamefont {G.}~\bibnamefont {Santos}}, \bibinfo {author} {\bibfnamefont {D.}~\bibnamefont {Juan}}, \bibinfo {author} {\bibfnamefont {S.~A.}\ \bibnamefont {Rosales}}, \bibinfo {author} {\bibfnamefont {J.}~\bibnamefont {Junquera}}, \bibinfo {author} {\bibfnamefont {P.}~\bibnamefont {Garc{\'\i}a-Fern{\'a}ndez}}, \bibinfo {author} {\bibfnamefont {S.}~\bibnamefont {Dicorato}}, \bibinfo {author} {\bibfnamefont {M.~M.}\ \bibnamefont {Giangregorio}}, \bibinfo {author} {\bibfnamefont {E.}~\bibnamefont {Dilonardo}}, \emph {et~al.},\ }\bibfield  {title} {\bibinfo {title} {Interlaboratory study on sb2s3 interplay between structure, dielectric function, and amorphous-to-crystalline phase change for photonics},\ }\href@noop {} {\bibfield  {journal} {\bibinfo  {journal} {Iscience}\ }\textbf {\bibinfo {volume} {25}} (\bibinfo {year}
  {2022})}\BibitemShut {NoStop}%
\bibitem [{\citenamefont {Barredo}\ \emph {et~al.}(2014)\citenamefont {Barredo}, \citenamefont {Parra}, \citenamefont {Guerrero}, \citenamefont {Fraile},\ and\ \citenamefont {Hermanns}}]{barredo2014mechanical}%
  \BibitemOpen
  \bibfield  {author} {\bibinfo {author} {\bibfnamefont {J.}~\bibnamefont {Barredo}}, \bibinfo {author} {\bibfnamefont {V.}~\bibnamefont {Parra}}, \bibinfo {author} {\bibfnamefont {I.}~\bibnamefont {Guerrero}}, \bibinfo {author} {\bibfnamefont {A.}~\bibnamefont {Fraile}},\ and\ \bibinfo {author} {\bibfnamefont {L.}~\bibnamefont {Hermanns}},\ }\bibfield  {title} {\bibinfo {title} {On the mechanical strength of monocrystalline, multicrystalline and quasi-monocrystalline silicon wafers: a four-line bending test study},\ }\href@noop {} {\bibfield  {journal} {\bibinfo  {journal} {Progress in Photovoltaics: Research and Applications}\ }\textbf {\bibinfo {volume} {22}},\ \bibinfo {pages} {1204} (\bibinfo {year} {2014})}\BibitemShut {NoStop}%
\bibitem [{\citenamefont {Fjellvåg}\ \emph {et~al.}(2023)\citenamefont {Fjellvåg}, \citenamefont {Waller}, \citenamefont {By},\ and\ \citenamefont {Sjastad}}]{fjellvåg2023pt}%
  \BibitemOpen
  \bibfield  {author} {\bibinfo {author} {\bibfnamefont {A.~S.}\ \bibnamefont {Fjellvåg}}, \bibinfo {author} {\bibfnamefont {D.}~\bibnamefont {Waller}}, \bibinfo {author} {\bibfnamefont {T.}~\bibnamefont {By}},\ and\ \bibinfo {author} {\bibfnamefont {A.~O.}\ \bibnamefont {Sjastad}},\ }\bibfield  {title} {\bibinfo {title} {Pt-catchment using pd/au alloys: Effect of enhanced diffusion},\ }\href@noop {} {\bibfield  {journal} {\bibinfo  {journal} {Industrial \& Engineering Chemistry Research}\ }\textbf {\bibinfo {volume} {62}},\ \bibinfo {pages} {2478} (\bibinfo {year} {2023})}\BibitemShut {NoStop}%
\bibitem [{\citenamefont {Jiang}\ \emph {et~al.}(2025)\citenamefont {Jiang}, \citenamefont {Zhang}, \citenamefont {Chen}, \citenamefont {He}, \citenamefont {Liu}, \citenamefont {Yu}, \citenamefont {Gao}, \citenamefont {Hong}, \citenamefont {Wang}, \citenamefont {Zhang} \emph {et~al.}}]{jiang2025two}%
  \BibitemOpen
  \bibfield  {author} {\bibinfo {author} {\bibfnamefont {H.}~\bibnamefont {Jiang}}, \bibinfo {author} {\bibfnamefont {X.}~\bibnamefont {Zhang}}, \bibinfo {author} {\bibfnamefont {K.}~\bibnamefont {Chen}}, \bibinfo {author} {\bibfnamefont {X.}~\bibnamefont {He}}, \bibinfo {author} {\bibfnamefont {Y.}~\bibnamefont {Liu}}, \bibinfo {author} {\bibfnamefont {H.}~\bibnamefont {Yu}}, \bibinfo {author} {\bibfnamefont {L.}~\bibnamefont {Gao}}, \bibinfo {author} {\bibfnamefont {M.}~\bibnamefont {Hong}}, \bibinfo {author} {\bibfnamefont {Y.}~\bibnamefont {Wang}}, \bibinfo {author} {\bibfnamefont {Z.}~\bibnamefont {Zhang}}, \emph {et~al.},\ }\bibfield  {title} {\bibinfo {title} {Two-dimensional czochralski growth of single-crystal mos2},\ }\href@noop {} {\bibfield  {journal} {\bibinfo  {journal} {Nature Materials}\ ,\ \bibinfo {pages} {1}} (\bibinfo {year} {2025})}\BibitemShut {NoStop}%
\bibitem [{\citenamefont {Tseng}\ \emph {et~al.}(2020)\citenamefont {Tseng}, \citenamefont {Jahani}, \citenamefont {Leitis},\ and\ \citenamefont {Altug}}]{tseng2020dielectric}%
  \BibitemOpen
  \bibfield  {author} {\bibinfo {author} {\bibfnamefont {M.~L.}\ \bibnamefont {Tseng}}, \bibinfo {author} {\bibfnamefont {Y.}~\bibnamefont {Jahani}}, \bibinfo {author} {\bibfnamefont {A.}~\bibnamefont {Leitis}},\ and\ \bibinfo {author} {\bibfnamefont {H.}~\bibnamefont {Altug}},\ }\bibfield  {title} {\bibinfo {title} {Dielectric metasurfaces enabling advanced optical biosensors},\ }\href@noop {} {\bibfield  {journal} {\bibinfo  {journal} {ACS photonics}\ }\textbf {\bibinfo {volume} {8}},\ \bibinfo {pages} {47} (\bibinfo {year} {2020})}\BibitemShut {NoStop}%
\bibitem [{\citenamefont {Huang}\ \emph {et~al.}(2020)\citenamefont {Huang}, \citenamefont {Zhang}, \citenamefont {Xiao}, \citenamefont {Wang}, \citenamefont {Fan}, \citenamefont {Liu}, \citenamefont {Zhang}, \citenamefont {Qu}, \citenamefont {Ji}, \citenamefont {Han} \emph {et~al.}}]{huang2020ultrafast}%
  \BibitemOpen
  \bibfield  {author} {\bibinfo {author} {\bibfnamefont {C.}~\bibnamefont {Huang}}, \bibinfo {author} {\bibfnamefont {C.}~\bibnamefont {Zhang}}, \bibinfo {author} {\bibfnamefont {S.}~\bibnamefont {Xiao}}, \bibinfo {author} {\bibfnamefont {Y.}~\bibnamefont {Wang}}, \bibinfo {author} {\bibfnamefont {Y.}~\bibnamefont {Fan}}, \bibinfo {author} {\bibfnamefont {Y.}~\bibnamefont {Liu}}, \bibinfo {author} {\bibfnamefont {N.}~\bibnamefont {Zhang}}, \bibinfo {author} {\bibfnamefont {G.}~\bibnamefont {Qu}}, \bibinfo {author} {\bibfnamefont {H.}~\bibnamefont {Ji}}, \bibinfo {author} {\bibfnamefont {J.}~\bibnamefont {Han}}, \emph {et~al.},\ }\bibfield  {title} {\bibinfo {title} {Ultrafast control of vortex microlasers},\ }\href@noop {} {\bibfield  {journal} {\bibinfo  {journal} {Science}\ }\textbf {\bibinfo {volume} {367}},\ \bibinfo {pages} {1018} (\bibinfo {year} {2020})}\BibitemShut {NoStop}%
\bibitem [{\citenamefont {Liu}\ \emph {et~al.}(2019)\citenamefont {Liu}, \citenamefont {Xu}, \citenamefont {Lin}, \citenamefont {Xiang}, \citenamefont {Feng}, \citenamefont {Cao}, \citenamefont {Li}, \citenamefont {Lan},\ and\ \citenamefont {Liu}}]{liu2019high}%
  \BibitemOpen
  \bibfield  {author} {\bibinfo {author} {\bibfnamefont {Z.}~\bibnamefont {Liu}}, \bibinfo {author} {\bibfnamefont {Y.}~\bibnamefont {Xu}}, \bibinfo {author} {\bibfnamefont {Y.}~\bibnamefont {Lin}}, \bibinfo {author} {\bibfnamefont {J.}~\bibnamefont {Xiang}}, \bibinfo {author} {\bibfnamefont {T.}~\bibnamefont {Feng}}, \bibinfo {author} {\bibfnamefont {Q.}~\bibnamefont {Cao}}, \bibinfo {author} {\bibfnamefont {J.}~\bibnamefont {Li}}, \bibinfo {author} {\bibfnamefont {S.}~\bibnamefont {Lan}},\ and\ \bibinfo {author} {\bibfnamefont {J.}~\bibnamefont {Liu}},\ }\bibfield  {title} {\bibinfo {title} {High-q quasibound states in the continuum for nonlinear metasurfaces},\ }\href@noop {} {\bibfield  {journal} {\bibinfo  {journal} {Physical review letters}\ }\textbf {\bibinfo {volume} {123}},\ \bibinfo {pages} {253901} (\bibinfo {year} {2019})}\BibitemShut {NoStop}%
\bibitem [{\citenamefont {Tian}\ \emph {et~al.}(2023)\citenamefont {Tian}, \citenamefont {Zhou}, \citenamefont {Abraham},\ and\ \citenamefont {Liu}}]{tian2023tunable}%
  \BibitemOpen
  \bibfield  {author} {\bibinfo {author} {\bibfnamefont {F.}~\bibnamefont {Tian}}, \bibinfo {author} {\bibfnamefont {J.}~\bibnamefont {Zhou}}, \bibinfo {author} {\bibfnamefont {E.}~\bibnamefont {Abraham}},\ and\ \bibinfo {author} {\bibfnamefont {Z.}~\bibnamefont {Liu}},\ }\bibfield  {title} {\bibinfo {title} {Tunable dielectric bic metasurface for high resolution optical filters},\ }\href@noop {} {\bibfield  {journal} {\bibinfo  {journal} {Journal of Physics D: Applied Physics}\ }\textbf {\bibinfo {volume} {56}},\ \bibinfo {pages} {134002} (\bibinfo {year} {2023})}\BibitemShut {NoStop}%
\bibitem [{\citenamefont {Zhang}\ \emph {et~al.}(2021)\citenamefont {Zhang}, \citenamefont {Fowler}, \citenamefont {Liang}, \citenamefont {Azhar}, \citenamefont {Shalaginov}, \citenamefont {Deckoff-Jones}, \citenamefont {An}, \citenamefont {Chou}, \citenamefont {Roberts}, \citenamefont {Liberman} \emph {et~al.}}]{zhang2021electrically}%
  \BibitemOpen
  \bibfield  {author} {\bibinfo {author} {\bibfnamefont {Y.}~\bibnamefont {Zhang}}, \bibinfo {author} {\bibfnamefont {C.}~\bibnamefont {Fowler}}, \bibinfo {author} {\bibfnamefont {J.}~\bibnamefont {Liang}}, \bibinfo {author} {\bibfnamefont {B.}~\bibnamefont {Azhar}}, \bibinfo {author} {\bibfnamefont {M.~Y.}\ \bibnamefont {Shalaginov}}, \bibinfo {author} {\bibfnamefont {S.}~\bibnamefont {Deckoff-Jones}}, \bibinfo {author} {\bibfnamefont {S.}~\bibnamefont {An}}, \bibinfo {author} {\bibfnamefont {J.~B.}\ \bibnamefont {Chou}}, \bibinfo {author} {\bibfnamefont {C.~M.}\ \bibnamefont {Roberts}}, \bibinfo {author} {\bibfnamefont {V.}~\bibnamefont {Liberman}}, \emph {et~al.},\ }\bibfield  {title} {\bibinfo {title} {Electrically reconfigurable non-volatile metasurface using low-loss optical phase-change material},\ }\href@noop {} {\bibfield  {journal} {\bibinfo  {journal} {Nature Nanotechnology}\ }\textbf {\bibinfo {volume} {16}},\ \bibinfo {pages} {661} (\bibinfo {year} {2021})}\BibitemShut {NoStop}%
\bibitem [{\citenamefont {Mikheeva}\ \emph {et~al.}(2019)\citenamefont {Mikheeva}, \citenamefont {Koshelev}, \citenamefont {Choi}, \citenamefont {Kruk}, \citenamefont {Lumeau}, \citenamefont {Abdeddaim}, \citenamefont {Voznyuk}, \citenamefont {Enoch},\ and\ \citenamefont {Kivshar}}]{mikheeva2019photosensitive}%
  \BibitemOpen
  \bibfield  {author} {\bibinfo {author} {\bibfnamefont {E.}~\bibnamefont {Mikheeva}}, \bibinfo {author} {\bibfnamefont {K.}~\bibnamefont {Koshelev}}, \bibinfo {author} {\bibfnamefont {D.-Y.}\ \bibnamefont {Choi}}, \bibinfo {author} {\bibfnamefont {S.}~\bibnamefont {Kruk}}, \bibinfo {author} {\bibfnamefont {J.}~\bibnamefont {Lumeau}}, \bibinfo {author} {\bibfnamefont {R.}~\bibnamefont {Abdeddaim}}, \bibinfo {author} {\bibfnamefont {I.}~\bibnamefont {Voznyuk}}, \bibinfo {author} {\bibfnamefont {S.}~\bibnamefont {Enoch}},\ and\ \bibinfo {author} {\bibfnamefont {Y.}~\bibnamefont {Kivshar}},\ }\bibfield  {title} {\bibinfo {title} {Photosensitive chalcogenide metasurfaces supporting bound states in the continuum},\ }\href@noop {} {\bibfield  {journal} {\bibinfo  {journal} {Optics express}\ }\textbf {\bibinfo {volume} {27}},\ \bibinfo {pages} {33847} (\bibinfo {year} {2019})}\BibitemShut {NoStop}%
\bibitem [{\citenamefont {Koshelev}\ \emph {et~al.}(2018)\citenamefont {Koshelev}, \citenamefont {Lepeshov}, \citenamefont {Liu}, \citenamefont {Bogdanov},\ and\ \citenamefont {Kivshar}}]{koshelev2018asymmetric}%
  \BibitemOpen
  \bibfield  {author} {\bibinfo {author} {\bibfnamefont {K.}~\bibnamefont {Koshelev}}, \bibinfo {author} {\bibfnamefont {S.}~\bibnamefont {Lepeshov}}, \bibinfo {author} {\bibfnamefont {M.}~\bibnamefont {Liu}}, \bibinfo {author} {\bibfnamefont {A.}~\bibnamefont {Bogdanov}},\ and\ \bibinfo {author} {\bibfnamefont {Y.}~\bibnamefont {Kivshar}},\ }\bibfield  {title} {\bibinfo {title} {Asymmetric metasurfaces with high-q resonances governed by bound states in the continuum},\ }\href@noop {} {\bibfield  {journal} {\bibinfo  {journal} {Physical review letters}\ }\textbf {\bibinfo {volume} {121}},\ \bibinfo {pages} {193903} (\bibinfo {year} {2018})}\BibitemShut {NoStop}%
\bibitem [{\citenamefont {Cueff}\ \emph {et~al.}(2024)\citenamefont {Cueff}, \citenamefont {Berguiga},\ and\ \citenamefont {Nguyen}}]{cueff2024fourier}%
  \BibitemOpen
  \bibfield  {author} {\bibinfo {author} {\bibfnamefont {S.}~\bibnamefont {Cueff}}, \bibinfo {author} {\bibfnamefont {L.}~\bibnamefont {Berguiga}},\ and\ \bibinfo {author} {\bibfnamefont {H.~S.}\ \bibnamefont {Nguyen}},\ }\bibfield  {title} {\bibinfo {title} {Fourier imaging for nanophotonics},\ }\href@noop {} {\bibfield  {journal} {\bibinfo  {journal} {Nanophotonics}\ }\textbf {\bibinfo {volume} {13}},\ \bibinfo {pages} {841} (\bibinfo {year} {2024})}\BibitemShut {NoStop}%
\end{thebibliography}%

\end{document}